\documentclass[amsmath,amssymb,showkeys,superscriptaddress,floatfix,prd]{revtex4}
\usepackage{graphicx,epstopdf}
\usepackage{subfig}
\usepackage{xcolor}
\usepackage[colorlinks=true, urlcolor=blue, linkcolor=blue, citecolor=green]{hyperref}
\def\pslash{p\!\!\!\slash }
\def\qslash{q\!\!\!\slash }
\def\xslash{x\!\!\!\slash }
\def\eslash{\varepsilon\!\!\!\slash }

\def\vel{\left|}
\def\ver{\right|}

\begin{document}

\title{Magnetic moments of doubly heavy baryons in light-cone QCD}

\author{Ula\c{s}~\"{O}zdem}%
\email[]{uozdem@dogus.edu.tr}
\affiliation{Department of Physics, Dogus University, Acibadem-Kadikoy, 34722 Istanbul, Turkey}

\date{\today}

\begin{abstract}
The magnetic dipole moments of  the spin-$\frac{1}{2}$  doubly heavy baryons are
extracted in the framework of light-cone QCD sum rule using the photon distribution amplitudes.
The electromagnetic properties of the doubly heavy baryons encodes important information of their internal structure and geometric shape.
The results for the magnetic dipole
moments of doubly heavy baryons acquired in this work are compared with the predictions
of the other theoretical approaches. The agreement of the estimations with some (but not all) theoretical estimations is good.
\end{abstract}

\maketitle

\section{Introduction}

A doubly charmed baryon was first reported by the SELEX
Collaboration in the decay mode $\Xi_{cc}^+ \rightarrow \Lambda_c^+ K^- \pi^+$
with the mass $M_{\Xi^+_{cc}}=3519 \pm 1~MeV$~\cite{Mattson:2002vu},
however, other experimental groups, namely Belle~\cite{Chistov:2006zj}, FOCUS~\cite{Ratti:2003ez},
 and BABAR~\cite{Aubert:2006qw} could not find any evidence of the doubly heavy baryons
 in $e^-\,e^+$ annihilations later.
 However, since the production mechanisms
at these experiments were different from that of SELEX Collaboration, which studied collisions of a hyperon
beam on fixed nuclear targets, these results had not ruled out the results of the SELEX Collaboration.
 %
 %
In 2017, LHCb Collaboration discovered  spin-$\frac{1}{2}$ doubly heavy baryon $\Xi_{cc}^{++}$ in the mass spectrum of
$\Lambda_c^+\,K^-\pi^+\,\pi^+ $ with the mass $M_{\Xi_{cc}^{++}}=3621.40\pm 0.72\pm0.27\pm0.14~MeV$~\cite{Aaij:2017ueg}.
The examination for the doubly heavy baryons (DHB) may provide us with important information our understanding the nonperturbative
QCD effects.
One of the several aspects which makes the
physics of DHB attractive is that the binding of a light quark and two heavy quarks
 ensures a unique point of view for dynamics of confinement.
Furthermore, the weak decays of DHB give an insight to the dynamics of singly heavy baryons.
%
%
Therefore, the masses~\cite{Bagan:1992za,Roncaglia:1995az,Ebert:1996ec,Tong:1999qs,Itoh:2000um,Gershtein:2000nx,
Kiselev:2001fw,Kiselev:2002iy,Narodetskii:2001bq,Lewis:2001iz,Ebert:2002ig,Mathur:2002ce,Flynn:2003vz,Vijande:2004at,
Chiu:2005zc,Migura:2006ep,Albertus:2006ya,Martynenko:2007je,Tang:2011fv,Liu:2007fg,Roberts:2007ni,
Valcarce:2008dr,Liu:2009jc,Alexandrou:2012xk,Aliev:2012ru,Aliev:2012iv,Namekawa:2013vu,Karliner:2014gca,Sun:2014aya,
Chen:2015kpa,Sun:2016wzh,Shah:2016vmd,Kiselev:2017eic,Chen:2017sbg,Hu:2005gf,Meng:2017fwb,
Narison:2010py,Zhang:2008rt,Guo:2017vcf,Lu:2017meb,Xiao:2017udy,Weng:2018mmf,Can:2013zpa,Branz:2010pq,Bose:1980vy,Patel:2008xs,SilvestreBrac:1996bg,Patel:2007gx,Gadaria:2016omw}, 
magnetic moments~\cite{Can:2013zpa,Branz:2010pq,Bose:1980vy,Patel:2008xs,
SilvestreBrac:1996bg,Patel:2007gx,Gadaria:2016omw,JuliaDiaz:2004vh,Faessler:2006ft,Can:2013tna,Li:2017cfz,Bernotas:2012nz,Lichtenberg:1976fi,Oh:1991ws,Simonis:2018rld,Liu:2018euh,Blin:2018pmj,Meng:2017dni,Dhir:2009ax},  
radiative~\cite{Li:2017pxa,Yu:2017zst,Lu:2017meb,Cui:2017udv}, strong~\cite{Hu:2005gf,Xiao:2017udy}  and
weak decays~\cite{Albertus:2006ya,Li:2017ndo,Wang:2017mqp,Wang:2017azm,Shi:2017dto} of the DHB have been studied extensively in literature in the framework of the lattice QCD, quark models, 
chiral perturbation theory (ChPT), potential models, QCD sum rules (QCDSR), light-cone QCD sum rules (LCSR), SU(3) flavor symmetry, heavy quark effective theory (HQET), nonperturbative string approach, Faddeev approach,  Feynman-Hellmann theorem, extended on-mass-shell renormalization scheme (EOMS), local diquark approach, perturbative QCD (PQCD), light front approach and extended chromomagnetic model. 
It is worth  mentioning that, various models lead to quite different predictions for the dynamic and static properties of DHB, which may be used to distinguish these models.

In order to find out the inner structure of the baryons in the nonperturbative regime
of QCD, the main challenges are the determination of
the statical and dynamical features of baryons
such as their coupling constants, magnetic dipole moments, masses and so on, both experimentally and  theoretically.
The magnetic dipole moment of hadrons is one of the most important quantities in examination of
their electromagnetic structure, and can provide valuable insight in understanding
the mechanism of strong interactions at low energies.
Obviously, specifying the magnetic dipole moment is an significant step in our understanding of the hadron
properties based on quark-gluon degrees of freedom.
The magnitude and sign of the dipole magnetic moment is provide important
information on structure, size and shape of hadrons.
The magnetic dipole moments of the spin-1/2  DHB have been studied
in different theoretical models and approaches
~\cite{SilvestreBrac:1996bg,Patel:2007gx,Gadaria:2016omw,JuliaDiaz:2004vh,Faessler:2006ft,Can:2013tna,Li:2017cfz,Bernotas:2012nz,Lichtenberg:1976fi,Oh:1991ws,Simonis:2018rld,Liu:2018euh,Blin:2018pmj}.

In this study, the magnetic dipole moments of spin-$\frac{1}{2}$ DHB
 (hereafter we will denote these states as $B_{QQ}$) are extracted
in the framework of the light-cone QCD sum rule (LCSR).
The LCSR has already been successfully applied to extract dynamical and
statical properties of hadrons for decades such as, form factors,
masses, the electromagnetic multipole
moments and so on. In this approach, the hadronic
features are expressed in terms of the features of the vacuum and the light cone distribution amplitudes (DAs) of the
hadrons in the process~[for details, see for instance~\cite{Chernyak:1990ag, Braun:1988qv, Balitsky:1989ry}].
 Since the magnetic dipole moment is expressed in terms of the properties of the DAs and the
QCD vacuum, any uncertainty in these parameters reflects the uncertainty of the estimations of
the magnetic dipole moments.

The rest of the paper is organized as follows:
In Sections II, the details of the magnetic dipole moments calculations
for the DHB with spin-$\frac{1}{2}$
 are presented.
In the last section, we numerically analyze the
sum rules obtained for the magnetic dipole moments and discuss
the obtained results.

\section{Formalism}

In this section  we derive the LCSR for the magnetic moments of spin-$\frac{1}{2}$ DHB.
The starting point is to consider the following correlation function:
\begin{equation}
 \label{edmn01}
\Pi(p,q)=i\int d^{4}xe^{ip\cdot x}\langle 0|\mathcal{T}\{J_{B_{QQ}}(x)
\bar J_{B_{QQ}}(0)\}|0\rangle_{\gamma},
\end{equation}%
where $\gamma$ is the external electromagnetic field
and $J_{B_{QQ}}$ is the interpolating current having quantum numbers $J^P =\frac{1}{2}^+$.
In this work, we choose the general form of the interpolating current for the spin-1/2 DHB as~\cite{Narison:2010py}
%
\begin{equation}
\label{cur1}
  J_{B_{QQ}}(x) = \varepsilon^{abc} \Big\{ \big(Q_a^T(x)C q_b(x)\big)\gamma_5 Q_c(x)
 +\beta \big[\big( Q_a^T(x)C\gamma_5 q_b(x)\big) Q_c(x)  \big]\Big\},
\end{equation}
where Q is the c or b-quark, q is u, d or s-quark,
C is the charge conjugation matrix; and a, b, and c are color indices
; and $\beta$
 is an arbitrary parameter that fixes the mixing of two local operators. 
 Choosing $\beta =-1$  makes the interpolating currents, which are known as Ioffe currents.

In order to obtain the sum rules for magnetic moments of the DHB the
above-mentioned correlation function is acquired from the following three steps:
\vspace*{0.3cm}

\textbullet ~being saturated by the hadrons having the same quantum numbers as the
interpolating currents (hadronic side),
\vspace*{0.3cm}

\textbullet ~ in terms of quark degrees of freedom interacting with the nonperturbative QCD vacuum
(QCD side).
\vspace*{0.3cm}

\textbullet ~ Then equating these two different representations of the correlation function to each other using the quark-hadron duality
assumption. In order to suppress the contributions of the higher states and continuum we carry out Borel transformation,
besides continuum subtraction to both sides of the acquired QCD sum rules.
\vspace*{0.3cm}

We start to calculate the correlation function in terms of hadronic degrees of freedom including the
physical properties of the baryons under consideration. To this end we insert intermediate states of $B_{QQ}$ into the
correlation function. As a result, we obtain
\begin{eqnarray}\label{edmn02}
\Pi^{Had}(p,q)&=&\frac{\langle0\mid J_{B_{QQ}}\mid
{B_{QQ}}(p)\rangle}{[p^{2}-m_{{B_{QQ}}}^{2}]}\langle {B_{QQ}}(p)\mid
{B_{QQ}}(p+q)\rangle_\gamma\frac{\langle {B_{QQ}}(p+q)\mid
\bar{J}_{B_{QQ}}\mid 0\rangle}{[(p+q)^{2}-m_{{B_{QQ}}}^{2}]}+...,
\end{eqnarray}
where q is the momentum of the photon and dots refers to contribution of the higher states and continuum.
The matrix elements in Eq.(\ref{edmn02}) are determined as
\begin{eqnarray}\label{edmn03}
\langle0\mid J_{B_{QQ}}(0)\mid {B_{QQ}}(p,s)\rangle&=&\lambda_{{B_{QQ}}}u(p,s),\\
\langle {B_{QQ}}(p)\mid {B_{QQ}}(p+q)\rangle_\gamma &=&\varepsilon^\mu\bar u(p)\Bigg[\big(f_1(q^2)+f_2(q^2)\big) \gamma_\mu
+f_2(q^2)\, \frac{(2p+q)_\mu}{2 m_{B_{QQ}}}\Bigg]u(p),
\end{eqnarray}
where $\lambda_{{B_{QQ}}}$ is the residue and u(p) is the Dirac spinor.
Summation over spins of $B_{QQ}$ baryon is applied as:
\begin{equation}
\label{edmn04}
 \sum_s u(p,s)\bar u(u,s)=\pslash+m_{B_{QQ}},
\end{equation}

Substituting Eqs. (\ref{edmn02})-(\ref{edmn04}) in Eq. (\ref{edmn01}) for hadronic side we get
\begin{align}
\label{edmn05}
\Pi^{Had}(p,q)=&\frac{\lambda^2_{B_{QQ}}}{[(p+q)^2-m^2_{B_{QQ}}][p^2-m^2_{B_{QQ}}]}
  \bigg[\Big(f_1(q^2)+f_2(q^2)\Big)\Big(2\,\pslash\eslash\pslash+\pslash\eslash\qslash+m_{B_{QQ}}\,\pslash\eslash
  +
  2\,m_{B_{QQ}}\,\eslash\pslash\nonumber\\
  &+m_{B_{QQ}}\,\eslash\qslash+m_{B_{QQ}}^2\eslash
  \Big)+ other~ structures~proportional~ with~ the~ f_2(q^2) \bigg].
\end{align}
At $q^2=0$, the magnetic dipole moment is defined in terms of $f_1(q^2=0)$ and $f_2(q^2=0)$
form factors in the following way;
\begin{equation}
\label{edmn06}
 \mu_{B_{QQ}}= f_1(q^2=0)+f_2(q^2=0).
\end{equation}
Among a number of different structures present in Eq. (\ref{edmn05}), 
we choose $\eslash\qslash$ which contains the magnetic moment form factor
$f_1+f_2$.
As a result, the hadronic side of the correlation can be written in terms of magnetic dipole moment of
the spin-$\frac{1}{2}$ DHB as,
\begin{eqnarray}
\label{edmn07}
\Pi^{Had}(p,q)&=&\mu_{B_{QQ}}\frac{\lambda^2_{B_{QQ}}\,m_{B_{QQ}}}{[(p+q)^2-m^2_{B_{QQ}}][p^2-m^2_{B_{QQ}}]}.
\end{eqnarray}

The next step is to compute the correlation function in terms of quark-gluon degrees of freedom
in the deep Euclidean region. Using the expression for interpolating current and Wick's theorem,
the QCD side of the correlation function can be written as,
\begin{align}
\label{edmn08}
 \Pi^{QCD}(p,q)&=i\, \varepsilon^{abc}\varepsilon^{a^{\prime}b^{\prime}c^{\prime}} \int d^4\,x e^{ip\cdot x}
 \langle 0\mid \Big\{
 \gamma_5 S_Q^{cc^{\prime}}(x)\gamma_5 Tr[\tilde S_Q^{aa^{\prime}}(x)S_q^{bb^{\prime}}(x)]
 -\gamma_5 S_Q^{ca^{\prime}}(x)\tilde S_q^{bb^{\prime}}(x)S_Q^{ac^{\prime}}(x)\gamma_5\nonumber\\
 &+\beta \Big(\gamma_5 S_Q^{cc^{\prime}}(x) Tr[\gamma_5 \tilde S_Q^{aa^{\prime}}(x)S_q^{bb^{\prime}}(x)]
 -\gamma_5 S_Q^{ca^{\prime}}(x)\gamma_5\tilde S_q^{bb^{\prime}}(x)S_Q^{ac^{\prime}}(x)
 +S_Q^{ac^{\prime}}(x)\gamma_5 Tr[\tilde S_Q^{ca^{\prime}}(x) \gamma_5 S_q^{bb^{\prime}}(x)]\nonumber\\
& -S_Q^{ac^{\prime}}(x) \tilde S_q^{bb^{\prime}}(x) \gamma_5 S_Q^{ca^{\prime}}(x) \gamma_5 \Big)
+\beta^2 \Big( S_Q^{cc^{\prime}}(x) Tr[\gamma_5 \tilde S_Q^{aa^{\prime}}(x) \gamma_5 S_q^{bb^{\prime}}(x) ]
-S_Q^{ca^{\prime}}(x) \gamma_5 \tilde S_q^{bb^{\prime}}(x)\gamma_5 S_Q^{ac^{\prime}}(x)\Big)
\Big\}\mid 0\rangle_\gamma
\end{align}
where $\tilde S_{Q(q)}^{ij}(x) = CS_{Q(q)}^{{ij}^T}(x)C$ and, $S_q^{ij}(x)$ and
$S_Q^{ij}(x)$ are the light and heavy quark propagators, respectively.
The light quark propagator is given as~\cite{Yang:1993bp},
\begin{eqnarray}
\label{edmn09}
S_{q}(x)=
\frac{1}{2 \pi^2 x^2}\Big( i \frac{{\xslash}}{x^{2}}-\frac{m_{q}}{2 } \Big)
- \frac{ \bar qq }{12} \Big(1-i\frac{m_{q} \xslash}{4}   \Big)
- \frac{\bar q \sigma.G q }{192}x^2  \Big(1-i\frac{m_{q} \xslash}{6}   \Big)
-\frac {i g_s }{32 \pi^2 x^2} ~G^{\mu \nu} (x) \bigg[\rlap/{x}
\sigma_{\mu \nu} +  \sigma_{\mu \nu} \rlap/{x}
 \bigg],
\end{eqnarray}%
where $G^{\mu \nu }$ is the gluon field strength tensor. The light quark propagator receives contributions
from non-local four-quark $\bar q q \bar q q$, three-particle $\bar q G q$, four-particle
$\bar q GGq$, etc. operators. In this study we consider contributions coming only from non-local operators with one gluon.
The contributions of four-quark and two-gluon-two-quark operators are omitted because of their small contributions.
The terms proportional to the four-quark and two-gluon-two-quark operators are neglected because of their small contribution.

The heavy quark propagator is given 
as~\cite{Belyaev:1985wza},
\begin{eqnarray}
\label{edmn10}
S_{Q}(x)&=&\frac{m_{Q}^{2}}{4 \pi^{2}} \bigg[ \frac{K_{1}(m_{Q}\sqrt{-x^{2}}) }{\sqrt{-x^{2}}}
+i\frac{{\xslash}~K_{2}( m_{Q}\sqrt{-x^{2}})}
{(\sqrt{-x^{2}})^{2}}\bigg]
-\frac{g_{s}m_{Q}}{16\pi ^{2}} \int_{0}^{1}dv~G^{\mu \nu }(vx)\bigg[ \big(\sigma _{\mu \nu }{\xslash}
  +{\xslash}\sigma _{\mu \nu }\big)\frac{K_{1}( m_{Q}\sqrt{-x^{2}}) }{\sqrt{-x^{2}}}\nonumber\\
&&+2\sigma ^{\mu \nu }K_{0}( m_{Q}\sqrt{-x^{2}})\bigg],
\end{eqnarray}%
where $K_{1,2}$ are Bessel functions of the second kind.

\begin{figure}[htp]
\subfloat[]{ \includegraphics[width=0.7\textwidth]{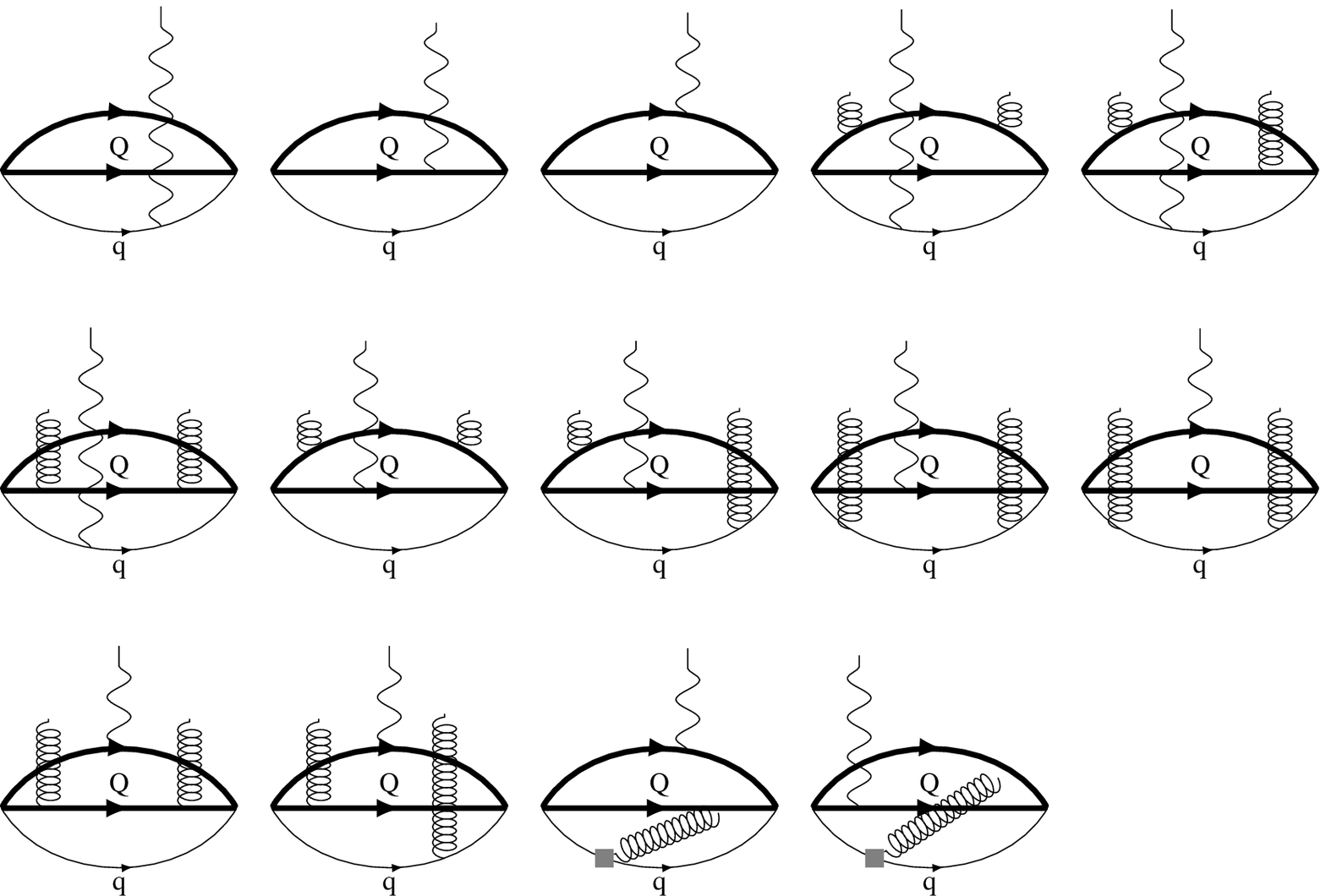}}\\
\subfloat[]{ \includegraphics[width=0.7\textwidth]{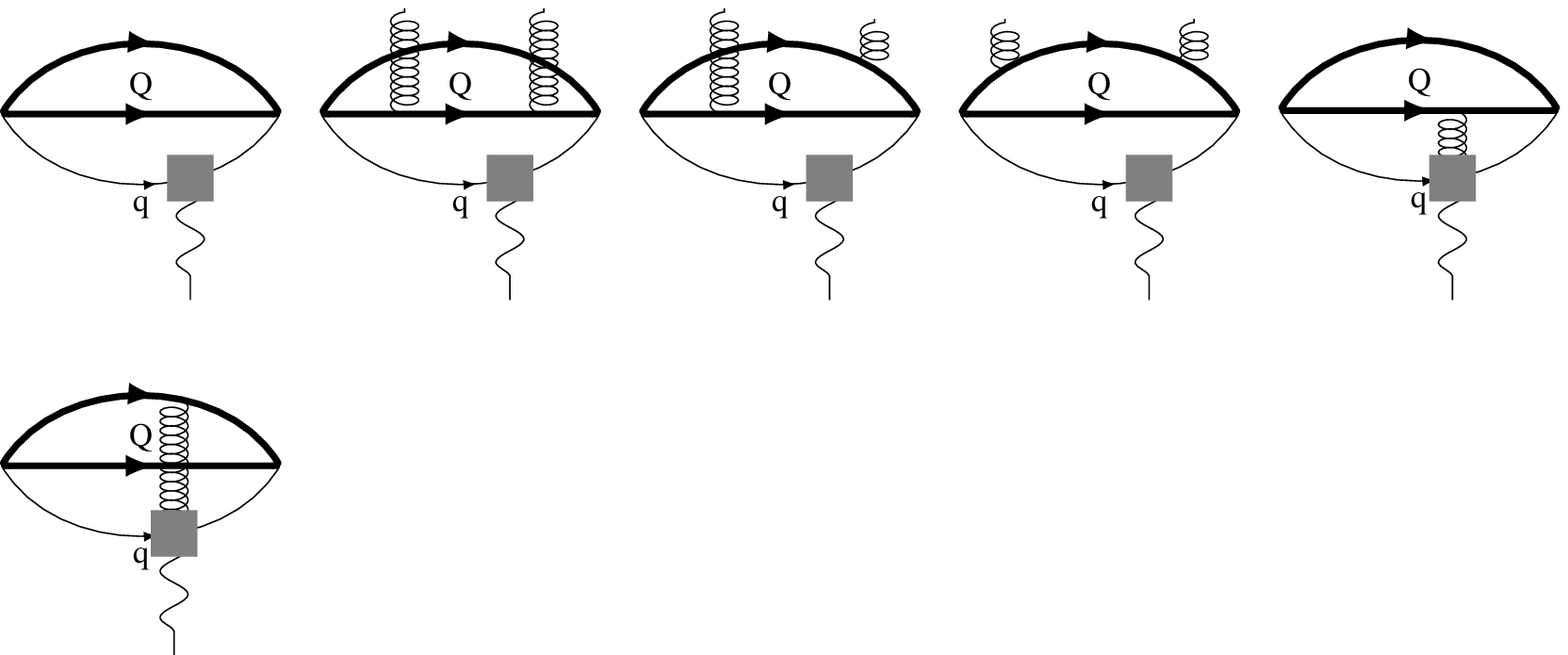}}
 \caption{ Feynman diagrams for the magnetic moments of the spin-1/2 DHB. The thick, thin, wavy and
curly lines represent the heavy quark, light quark, photon and gluon propagators, respectively. 
 Diagrams (a) corresponding to the perturbative photon vertex and,  
 diagrams (b) represent the contributions coming from the DAs of the photon.}
  \end{figure}

 The correlation function in Eq. (\ref{edmn08}) includes different contributions: the photon can be emitted both perturbatively or nonperturbatively.
 When the photon is emitted, perturbatively, one of the propagators in Eq. (\ref{edmn08}) is replaced by
 \begin{eqnarray}
\label{edmn12}
S^{free} (x) \rightarrow \int d^4y\, S^{free} (x-y)\,\rlap/{\!A}(y)\, S^{free} (y)\,,
\end{eqnarray}
where $S^{free}$ is the first term of the light or heavy quark propagators and the remaining propagators are replaced with the
full quark propagators including the free (perturbative) part as well as the
interacting parts (with gluon or QCD vacuum) as nonperturbative contributions.
Here we use $ A_\alpha(y)=-\frac{1}{2}\, F_{\alpha\beta}(y)\, y^\beta $ where  the electromagnetic field strength tensor is written as $ F_{\alpha\beta}(y)=-i(\varepsilon_\alpha q_\beta-\varepsilon_\beta q_\alpha)\,e^{iq.y} $.
The total perturbative photon emission is acquired by performing the
replacement mentioned above for the perturbatively interacting quark propagator with the photon and
making use of the replacement of the remaining propagators by their free parts.

In case of nonperturbative photon emission the light quark propagator in Eq. (\ref{edmn08}) is replaced by
\begin{align}
\label{edmn13}
S_{\alpha\beta}^{ab} \rightarrow -\frac{1}{4} (\bar{q}^a \Gamma_i q^b)(\Gamma_i)_{\alpha\beta},
\end{align}
and remaining two propagators are replaced with the full quark propagators, and also including perturbative
and nonperturbative contributions.
 Once Eq. (\ref{edmn13}) is inserted into Eq. (\ref{edmn08}), there seem matrix
elements such as $\langle \gamma(q)\vel \bar{q}(x) \Gamma_i q(0) \ver 0\rangle$
and $\langle \gamma(q)\vel \bar{q}(x) \Gamma_i G_{\alpha\beta}q(0) \ver 0\rangle$,
representing the nonperturbative contributions.
To compute the nonperturbative contributions,
we need the matrix elements of the nonlocal operators between the vacuum
and the photon states and
these matrix elements are defined in terms of the photon DAs with definite
twists, whose expressions are given in the Appendix. 
In principle, the photon can be emitted at long distance from the heavy quarks. But due to the large mass of the heavy quarks, such long distance photon emission from the heavy quarks will be largely suppressed. Such contributions are ignored in our calculation. 
Only the short distance  photon emission from the heavy quarks are considered, as described in Eq. (\ref{edmn12}). 
For this reason, DAs involving the heavy quarks are not used in our calculation.
The QCD side of the correlation function
can be acquired in terms of quark and gluons degrees of freedom
by substituting photon DAs and expressions for heavy
and light quarks propagators in to Eq. (\ref{edmn08}).

The QCD sum rules for the magnetic dipole moments of the spin-$\frac{1}{2}$ DHB are obtained
by equating the coefficients of the structure $\eslash\qslash$ from hadronic and QCD sides of the correlation
function. The last step in deriving the sum rules for the
magnetic dipole moments of the spin-$\frac{1}{2}$ DHB are applying double Borel transformations
over the $p^2$ and $(p+q)^2$ on the both sides of the correlation function
in order to suppress the contributions of higher states and continuum.
Finally, we obtain;
\begin{eqnarray}
\label{edmn14}
 \mu_{B_{QQ}}=\frac{e^{\frac{m^2_{B_{QQ}}}{M^2}}}{\lambda^2_{B_{QQ}}\, m_{B_{QQ}}}\, \Pi^{QCD}.
\end{eqnarray}
The explicit forms of the $\Pi^{QCD}$ is given as follows:
\begin{align}
\Pi^{QCD}=& -\frac {3} {2621440\, \pi^3} \bigg[\bigg ( e_Q\,
      m_q \big(1 - \beta\big)^2 + e_q\,
      m_Q \big(1 - \beta^2\big)  \bigg) \bigg (I[0, 5, 2] - 3 I[0, 5, 3] + 3 I[0, 5, 4] - I[0, 5, 5]\bigg)\bigg]\nonumber\\
&-\frac { m_Q\, \langle g_s^2 G^2 \rangle } {3538944\, \pi^2} (e_q + 7\, e_Q)\big (1 - \beta^2 \big)   \bigg [I[0, 3, 1] - 2\,
  I[0, 3, 2] + I[0, 3, 3]\bigg]\nonumber\\
  &-\frac {16\,e_Q\, m_Q^2\, \langle \bar qq \rangle} {393216 \, \pi}\bigg[
   \big (3 + 2\, \beta + 3\, \beta^2\big)\, \Big (I[0, 3, 1] - 2 I[0, 3, 2] + I[0, 3, 3]\Big)\bigg]\nonumber\\
              &+\frac {e_q\, m_Q^2\, \langle \bar qq \rangle} {393216 \, \pi} \bigg[\big (1 + \beta\big)^2  \Big (2 I[0, 3, 2] - 4 I[0, 3, 3] + 2 I[0, 3, 4] +
        3 I[1, 2, 2] - 6 I[1, 2, 3] + 3 I[1, 2, 4]\Big) \mathbb A[u_0]\nonumber\\
        &+  4\, \big (1 - \beta \big)^2 \Big (I[0, 3, 1] - 2 I[0, 3, 2] + I[0, 3, 3]\Big) I_ 3[\mathcal S] + (1 + \beta)^2 \bigg\{-2 \
\Big (I[0, 3, 1] - 2 I[0, 3, 2] + I[0, 3, 3]\Big)\nonumber\\
&\times \Big (2 I_ 2[\mathcal T_ 1] +
           2 I_ 2[\mathcal T_ 2] - I_ 2[\mathcal T_ 3] -
           I_ 2[\mathcal T_ 4] - 2 I_ 2[\mathcal {\tilde S}]\Big) -
        2 \Big (I[0, 3, 2] - 2 I[0, 3, 3] + I[0, 3, 4]\Big) I_ 3[
          h_\gamma] + \chi \Big (I[0, 4, 2] \nonumber\\
          &- 2 I[0, 4, 3] +
            I[0, 4, 4]\Big) \varphi_\gamma[u_ 0]\bigg\}\bigg]\nonumber\\
&- \frac {12\,e_Q\, m_q\, m_Q \langle \bar qq \rangle} {393216 \, \pi}\bigg[ (1 - \beta^2)
\Big (6 I[0, 3, 1] - 19 I[0, 3, 2] + 20  I[0, 3, 3] - 7 I[0, 3, 4] + 9 I[1, 2, 1]- 28 I[1, 2, 2]\nonumber\\
  & + 29 I[1, 2, 3] - 10\,   I[1, 2, 4]\Big)\bigg] \nonumber\\
  &+\frac {e_q \langle \bar qq \rangle } {5242880\, \pi}\big (
    1 - \beta\big)^2 \bigg[5\Big (I[0, 4, 3] - 3 I[0, 4, 4] + 3 I[0, 4, 5] - I[0, 4, 6] +
    4 I[1, 3, 3] - 12 I[1, 3, 4] + 12 I[1, 3, 5] \nonumber\\
    &- 4 I[1, 3, 6] +  2 I[2, 2, 3] - 6 I[2, 2, 4] + 6 I[2, 2, 5] -
    2 I[2, 2, 6]\Big)\, \mathbb A[u_ 0]
    -5 \bigg \{\Big (I[0, 4, 2] - 3 I[0, 4, 3] + 3 I[0, 4, 4]\nonumber\\
    &- I[0, 4, 5] + 2 I[1, 3, 2] - 6 I[1, 3, 3] + 6 I[1, 3, 4] -
       2 I[1, 3, 5]\Big) \Big (I_ 2[\mathcal S] -
      2 I_ 2[\mathcal T_ 2] - I_ 2[\mathcal T_ 3] +
      I_ 2[\mathcal T_ 4]\Big)+ \Big (I[0, 4, 2]
    \nonumber\\
    &- 3 I[0, 4, 3] + 3 I[0, 4, 4] -  I[0, 4, 5]\Big)\Big (
      I_ 2[\mathcal T_ 1] + I_ 2[\mathcal {\tilde S}]\Big)\bigg\}
      +15 \Big (I[0, 4, 3] - 3 I[0, 4, 4] + 3 I[0, 4, 5]- I[0, 4, 6]  \nonumber\\
       &+     2 I[1, 3, 3]- 6 I[1, 3, 4] + 6 I[1, 3, 5] - 2 I[1, 3, 6]\Big)\, I_3[h_\gamma] -
 3\, \chi \, \Big (2 I[0, 5, 3] - 6 I[0, 5, 4] + 6 I[0, 5, 5]  \nonumber\\
  & -     2 I[0, 5, 6]+ 5 I[1, 4, 3]- 15 I[1, 4, 4] + 15 I[1, 4, 5] -
    5 I[1, 4, 6]\Big) \varphi_\gamma[u_ 0]
    \bigg]\nonumber\\
 &-\frac {3\,e_Q \langle \bar qq \rangle} {393216 \, \pi}\big (1 - \beta\big)^2\bigg[
    3\,  \Big (I[0, 4, 2] - 3 I[0, 4, 3]+ 3  I[0, 4, 4] - I[0, 4, 5]\Big)
  +4\, \Big (I[1, 3, 2] - 3 I[1, 3, 3]\nonumber\\
  &+ 3 I[1, 3, 4] - I[1, 3, 5]\Big)\bigg]\nonumber\\
 &+\frac {e_q\, m_Q \,f_ {3\gamma} } {3145728 \, \pi} \bigg[
     6 \big (1 + 4 \beta - \beta^2\big) \Big(I[0, 4, 1] - 3 I[0, 4, 2] +
         3 I[0, 4, 3] -
         I[0, 4, 4]\Big)\, I_ 1[\mathcal A] + \big (1 - \beta^2\big) \bigg \{\Big (12 I[0, 4, 1] \nonumber\\
         &- 37 I[0, 4, 2] + 38 I[0, 4, 3] -
             13 I[0, 4, 4]\Big) I_ 1[\mathcal V] +
          12\Big (I[0, 4, 2] - 3 I[0, 4, 3] + 3 I[0, 4, 4] -
              I[0, 4, 5]\Big) \psi^\nu[u_ 0]\bigg\}\bigg],
                         \end{align}
where 
\begin{align*}
 {M^2}= \frac{M_1^2 M_2^2}{M_1^2+M_2^2}, ~~~
 u_0= \frac{M_1^2}{M_1^2+M_2^2},
\end{align*}
%
with $ M_1^2 $ and $ M_2^2 $ being the Borel parameters in the initial and final states, respectively.
Since we have the same DHB in the initial and final states, therefore we can set, M$_1^2$ = M$_2^2$= 2 M$^2$, which leads to $u_0 = 1/2 $.
Physically, this means that each quark and antiquark carries half of the photon’s momentum.
 Here $m_Q$ is the mass of the c or b-quark, $m_q$ is the mass of the u, d or s-quark,
$e_Q$ is the electric charge of the c or b-quark,
$e_q$ is the electric charge of the u, d or s-quark,
$\chi$ is the magnetic susceptibility of the quark condensate,
$\langle \bar qq \rangle$ and $\langle g_s^2 G^2 \rangle $ are the quark
and gluon condensates, respectively.
%
%
The reader can find out some details about the computations
such as Borel transformations and continuum subtraction in Refs.~\cite{Agaev:2016srl,Azizi:2018duk}.


The functions~$I[n,m,l]$, $I_1[f]$,~$I_2[f]$ and ~$I_3[f]$ are defined as:
\begin{align}
 I[n,m,l]&= \int_{4 m_Q^2}^{s_0} ds \int_{0}^1 dt ~ e^{-s/M^2}~ s^n\,(s-4\,m_Q^2)^m\,t^l,\nonumber\\
 I_1[f]&=\int D_{\alpha_i} \int_0^1 dv~ f(\alpha_{\bar q},\alpha_q,\alpha_g)\, \delta'(\alpha_ q +(1-v) \alpha_g-u_0),\nonumber\\
  I_2[f]&=\int D_{\alpha_i} \int_0^1 dv~ f(\alpha_{\bar q},\alpha_q,\alpha_g)\, \delta(\alpha_ q +(1-v) \alpha_g-u_0),\nonumber\\
 I_3[f]&=\int_0^1 du~ f(u),
 \end{align}
where, $f$ represents the corresponding photon DAs, whose expressions are presented in the Appendix.

\section{Numerical analysis and conclusion}
In this section, we perform numerical analysis for
the spin-$\frac{1}{2}$ DHB.
We use  
$m_u =m_d =0$,
$m_s=96^{+8}_{-4}$~MeV,
$m_c = 1.28 \pm 0.03$~GeV, $m_b=4.18^{+0.04}_{-0.03}$~GeV,~\cite{Patrignani:2016xqp},
$M_{\Xi^+_{cc}}=3519 \pm 1$~MeV~\cite{Mattson:2002vu},
$M_{\Xi_{cc}^{++}}=3621.40\pm 1.13$~MeV~\cite{Aaij:2017ueg},
$f_{3\gamma}=-0.0039$~GeV$^2$~\cite{Ball:2002ps},
$\langle \bar qq\rangle =(-0.24\pm0.01)^3$~GeV$^3$ \cite{Ioffe:2005ym},
$m_0^{2} = 0.8 \pm 0.12$~GeV$^2$ 
 and $\chi=-2.85 \pm 0.5$~GeV$^2$~\cite{Rohrwild:2007yt}.
The masses of the $\Omega^+_{cc}$, $\Xi^{0}_{bb}$, $\Xi^{-}_{bb}$ and $\Omega^{-}_{bb}$
 baryons are borrowed from Ref.~\cite{Aliev:2012ru,Aliev:2012iv},
in which the mass sum rules have been used in computing them.
These masses are computed to have the following values:  $M_{\Omega^+_{cc}}= 3.73 \pm 0.2$~GeV,
$M_{\Xi^{0}_{bb}}=9.96 \pm 0.90$~GeV,
$M_{\Xi^{-}_{bb}}=9.96 \pm 0.90$~GeV, $M_{\Omega^{-}_{bb}}=9.97 \pm 0.90$~GeV.
In order to specify the magnetic dipole moments of DHB, the value of
the residues are needed. The residues of the DHB are computed in Refs.~\cite{Aliev:2012ru,Aliev:2012iv}.
These residues are calculated to have the following values:
$\lambda_{\Xi_{cc}}=0.16 \pm 0.03$~GeV$^3$, $\lambda_{\Omega_{cc}}=0.18 \pm 0.04$~GeV$^3$,
$\lambda_{\Xi_{bb}}=0.44 \pm 0.08$~GeV$^3$ and $\lambda_{\Omega_{bb}}=0.45 \pm 0.08$~GeV$^3$.
The parameters used in the photon DAs are given in the Appendix. 

The sum rules for the magnetic dipole moments of the DHB
 depend on three auxiliary parameters, namely the continuum threshold $s_0$, Borel mass parameter $M^2$
  and mixing parameter $\beta$.
 We shall find their working region such that the magnetic dipole moments are practically
 independent of these parameters according to the standard prescriptions
in QCD sum rules.
The continuum threshold is not totally arbitrary,
 it is chosen as the point at which the excited states and continuum begin to contribute to the computations.
 However, since we have very limited knowledge on the energy of excited states we should decide how to choose working interval of the continuum threshold.
To designate the working region of the $s_0$, we enforce the conditions of  OPE convergence and pole dominance.
 In this respect, we choose the value of the continuum threshold within the interval
 $s_0 = (16-20)$~GeV$^2$ for $\Xi_{cc}$, $s_0 = (18-22)$~GeV$^2$ for $\Omega_{cc}$,
 $s_0 = (116-120)$~GeV$^2$ for $\Xi_{bb}$ and $s_0 = (118-122)$~GeV$^2$ for $\Omega_{bb}$ baryons.
The working window for $M^2$ is acquired by requiring that the series of OPE in QCD side
is convergent and the contribution
of higher states and continuum is adequately suppressed.
Our numerical analysis shows that these conditions are fulfilled
when $M^2$ change in the regions: 4~GeV$^2$ $\leq$ M$^2$ $\leq$ 6~GeV$^2$ for $\Xi_{cc}$,
5~GeV$^2 \leq$ M$^2$ $\leq$ 7~GeV$^2$ for $\Omega_{cc}$, 10~GeV$^2 \leq$ M$^2 \leq$ 14~GeV$^2$ for $\Xi_{bb}$ and
11~GeV$^2 \leq$ M$^2 \leq$ 15~GeV$^2$ for $\Omega_{bb}$ baryons.
 In Fig. 2, we plot the dependencies of the magnetic dipole moments on $M^2$
 at several fixed values of the continuum threshold $s_0$.
 We observe from the figure that the magnetic dipole moments
 show relatively weak dependence on the variations of the Borel mass parameter
 and continuum threshold in their working regions.
The sum rules are expected to be independent of the
mixing parameter $\beta$ and it is chosen $\beta =\pm 2$~\cite{Aliev:2012ru}.

Our final results for the magnetic dipole moments are given in Table I.
For comparison, in the same Table we present the estimations of other approaches on the
magnetic dipole moments of the  spin-1/2 DHB.
The errors in the given results originate because of the variations in the  calculations of the working regions of
 $M^2$ and $s_0$ as well as the uncertainties
in the values of the input parameters and the photon DAs.
We should also stress that the primary
source of uncertainties is the variations with respect to $s_0$ and the results weakly depend on the
choices of the Borel mass parameter.

 \begin{table}[t]
\centering
\begin{tabular}{l|c|c|c|c|c|cccc}
\hline\noalign{\smallskip}
Approaches & $\Xi_{ccu}^{++}$ & $\Xi_{ccd}^{+}$ & $\Omega_{ccs}^{+}$&$\Xi_{bbu}^{0}$&$\Xi_{bbd}^{-}$&$\Omega_{bbs}^{-}$\\
\noalign{\smallskip}\hline\noalign{\smallskip}
QM~\cite{Lichtenberg:1976fi} & -0.12 & 0.80 & 0.69&-&-&-\\
RQM \cite{JuliaDiaz:2004vh} & -0.10 & 0.86 & 0.72&-&-&-\\
Skyrmion \cite{Oh:1991ws} & -0.47 & 0.98 & 0.59&-&-&-\\
NQM \cite{Patel:2007gx} &-0.20 & 0.79 & 0.64&-&-&-\\
Lattice QCD \cite{Can:2013tna} &- & 0.425 & 0.413&-&-&-\\
EOMS BHCPT-I\cite{Liu:2018euh} &- & 0.392 & 0.397&-&-&-\\
EOMS BHCPT-II \cite{Blin:2018pmj} &- & 0.37 & 0.40&-&-&-\\
MIT Bag model-I~\cite{Bernotas:2012nz} & 0.11 & 0.72 &  0.66&-0.43&0.09&0.04 \\
MIT Bag model~II\cite{Simonis:2018rld} & -0.11 & 0.72 & 0.64&-0.58&0.17&0.11 \\
RTQM \cite{Faessler:2006ft} & 0.13 & 0.72 & 0.67&-0.53&0.18&0.04\\
NRQM \cite{SilvestreBrac:1996bg} & -0.20 & 0.78 & 0.63&-0.69&0.23&0.10\\
RHM~\cite{Gadaria:2016omw}& -0.17 & 0.85 &0.74&-0.89&0.32&0.16\\
HBChBT~\cite{Li:2017cfz} & -0.25 & 0.85 &0.78&-0.84&0.26&0.19\\
This work&$-0.23\pm 0.05$&$0.43\pm 0.09$&$0.39\pm0.09$&$-0.51\pm 0.09$&$0.28 \pm 0.04$&$0.42\pm 0.05$\\
\noalign{\smallskip}\hline
\end{tabular}
\caption{Magnetic dipole moments of the spin-$\frac{1}{2}$ DHB (in units of $\mu_{N}$).}
 \end{table}

 In Table I, we compare our predictions with the results obtained
using other approaches, such as quark model
(QM)~\cite{Lichtenberg:1976fi}, relativistic three-quark model
(RTQM)~\cite{Faessler:2006ft}, nonrelativistic quark model in
Faddeev approach (NRQM)~\cite{SilvestreBrac:1996bg}, relativistic
quark model (RQM)~\cite{JuliaDiaz:2004vh}, skyrmion
model~\cite{Oh:1991ws},  MIT bag model~\cite{Bose:1980vy,Bernotas:2012nz,Simonis:2018rld},
nonrelativistic quark model (NQM)~\cite{Patel:2007gx},
relativistic harmonic confinement model (RHM)~\cite{Gadaria:2016omw}, lattice
QCD~\cite{Can:2013tna}, heavy baryon chiral perturbation theory (HBChBT)~\cite{Li:2017cfz}
 and covariant baryon chiral perturbation theory with the extended on-mass-shell
scheme (EOMS BHCPT)~\cite{Liu:2018euh,Blin:2018pmj}.
From a comparison of our results with the estimations of other approaches
we see that for the $\Xi_{ccu}^{++}$  baryon, all results more or less consistent with each other
except the results of Ref.~\cite{Oh:1991ws}, which is quite different.
For the $\Xi_{ccd}^{+}$ and  $\Omega_{ccs}^{+}$ baryons, 
consistent with Refs.~\cite{Can:2013tna,Liu:2018euh,Blin:2018pmj} and approximately two times smaller than other predictions.
For the $\Xi_{bbu}^{0}$  baryon we see that all results more or less, are similar
except the results of Ref.~\cite{Li:2017cfz,Gadaria:2016omw,SilvestreBrac:1996bg}, which are large.
For the $\Xi_{bbd}^{-}$ baryon, all results more or less consistent with each other
except the results of Ref.~\cite{Bernotas:2012nz}, which are small.
For the $\Omega_{bbs}^{-}$ baryon, our predictions are larger than other predictions.
As can be seen from this Table, various models lead to quite different predictions for
the magnetic dipole moments of DHB, which may be used to distinguish these models.
However, the direct measurement of the magnetic dipole moments of DHB are unlikely in the
near future. Hence, any indirect predictions of the magnetic dipole moments of the DHB could be very useful.

In conclusion, we have calculated the magnetic dipole moments
 of the spin-1/2 DHB in the
framework of light-cone QCD sum rule.
The electromagnetic properties of the DHB encodes
important information of their internal structure and geometric shape.
We performed a comparison of our results with the estimations of
various theoretical approaches existing in literature.
The agreement of the estimations with some (but not all) theoretical
estimations is good.
We hope our analysis may be helpful for future experimental measurements.
\begin{figure}
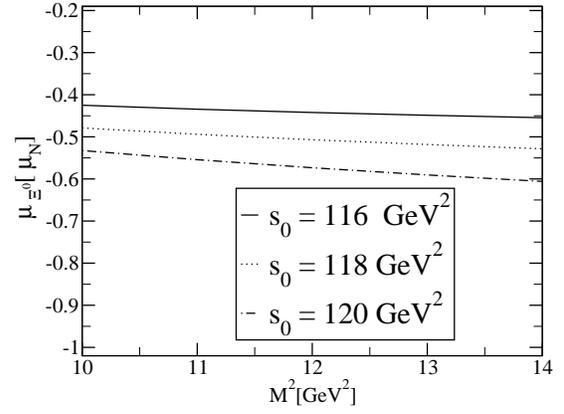
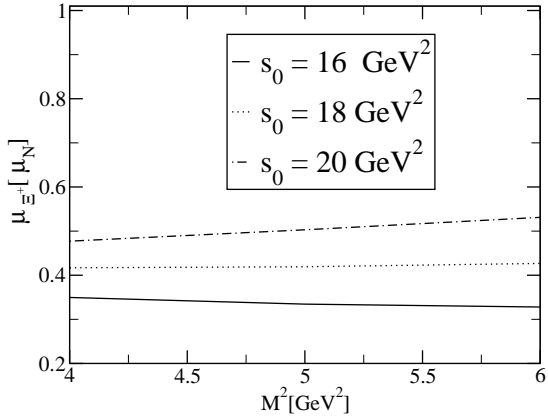
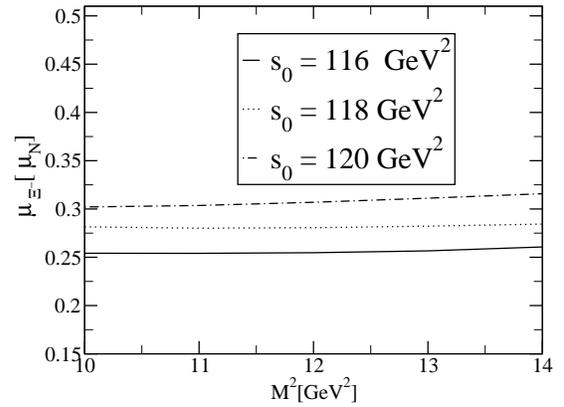
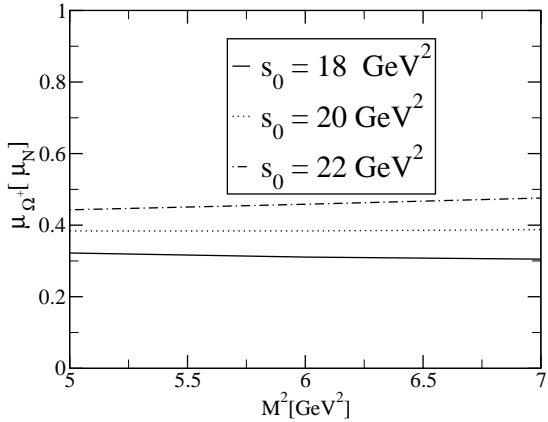
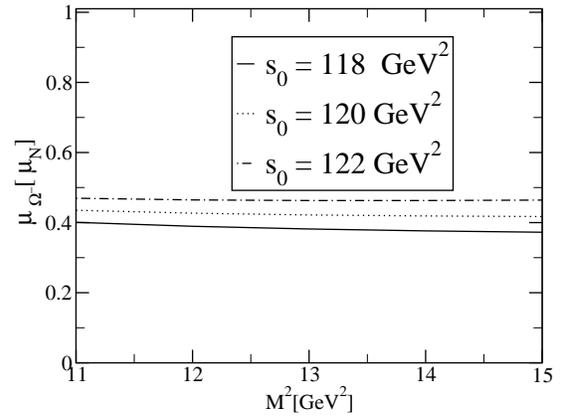

\centering
 \subfloat[]{\includegraphics[width=0.4\textwidth]{1XccuS-one.eps}}~~~~~~~~
 \subfloat[]{ \includegraphics[width=0.4\textwidth]{1XbbuS-one.eps}}\\
  \vspace{0.5cm}
 \subfloat[]{\includegraphics[width=0.4\textwidth]{1XccdS-one.eps}}~~~~~~~~
 \subfloat[]{ \includegraphics[width=0.4\textwidth]{1XbbdS-one.eps}}\\
  \vspace{0.5cm}
 \subfloat[]{\includegraphics[width=0.4\textwidth]{1OccsS-one.eps}}~~~~~~~~
 \subfloat[]{ \includegraphics[width=0.4\textwidth]{1ObbsS-one.eps}}
 \caption{ The dependence of the magnetic dipole moments for spin-$\frac{1}{2}$ DHB
 on the Borel parameter squared $M^{2}$
 at different fixed values of the continuum threshold:
 (a), (c) and (e) for the doubly charmed baryons,
 (b), (d) and (f) for the doubly bottomed baryons.}
  \end{figure}
\section{Acknowledgements}
The author is grateful to K. Azizi, A. Ozpineci and V. S. Zamiralov for helpful discussions.
\newpage
\appendix
\section*{Appendix: Photon distribution amplitudes}
		        In this appendix, we present the definitions of the matrix elements of the
form $\langle \gamma(q)\vel \bar{q}(x) \Gamma_i q(0) \ver 0\rangle$
and $\langle \gamma(q)\vel \bar{q}(x) \Gamma_i G_{\mu\nu}q(0) \ver 0\rangle$ in terms of the photon
DAs, and the explicit expressions of the
photon distribution amplitudes \cite{Ball:2002ps},
\begin{eqnarray*}
\label{esbs14}
&&\langle \gamma(q) \vert  \bar q(x) \gamma_\mu q(0) \vert 0 \rangle
= e_q f_{3 \gamma} \left(\varepsilon_\mu - q_\mu \frac{\varepsilon
x}{q x} \right) \int_0^1 du e^{i \bar u q x} \psi^v(u)
\nonumber \\
&&\langle \gamma(q) \vert \bar q(x) \gamma_\mu \gamma_5 q(0) \vert 0
\rangle  = - \frac{1}{4} e_q f_{3 \gamma} \epsilon_{\mu \nu \alpha
\beta } \varepsilon^\nu q^\alpha x^\beta \int_0^1 du e^{i \bar u q
x} \psi^a(u)
\nonumber \\
&&\langle \gamma(q) \vert  \bar q(x) \sigma_{\mu \nu} q(0) \vert  0
\rangle  = -i e_q \langle \bar q q \rangle (\varepsilon_\mu q_\nu - \varepsilon_\nu
q_\mu) \int_0^1 du e^{i \bar u qx} \left(\chi \varphi_\gamma(u) +
\frac{x^2}{16} \mathbb{A}  (u) \right) \nonumber \\
&&-\frac{i}{2(qx)}  e_q \bar qq \left[x_\nu \left(\varepsilon_\mu - q_\mu
\frac{\varepsilon x}{qx}\right) - x_\mu \left(\varepsilon_\nu -
q_\nu \frac{\varepsilon x}{q x}\right) \right] \int_0^1 du e^{i \bar
u q x} h_\gamma(u)
\nonumber \\
&&\langle \gamma(q) | \bar q(x) g_s G_{\mu \nu} (v x) q(0) \vert 0
\rangle = -i e_q \langle \bar q q \rangle \left(\varepsilon_\mu q_\nu - \varepsilon_\nu
q_\mu \right) \int {\cal D}\alpha_i e^{i (\alpha_{\bar q} + v
\alpha_g) q x} {\cal S}(\alpha_i)
\nonumber \\
&&\langle \gamma(q) | \bar q(x) g_s \tilde G_{\mu \nu}(v
x) i \gamma_5  q(0) \vert 0 \rangle = -i e_q \langle \bar q q \rangle \left(\varepsilon_\mu q_\nu -
\varepsilon_\nu q_\mu \right) \int {\cal D}\alpha_i e^{i
(\alpha_{\bar q} + v \alpha_g) q x} \tilde {\cal S}(\alpha_i)
\nonumber \\
&&\langle \gamma(q) \vert \bar q(x) g_s \tilde G_{\mu \nu}(v x)
\gamma_\alpha \gamma_5 q(0) \vert 0 \rangle = e_q f_{3 \gamma}
q_\alpha (\varepsilon_\mu q_\nu - \varepsilon_\nu q_\mu) \int {\cal
D}\alpha_i e^{i (\alpha_{\bar q} + v \alpha_g) q x} {\cal
A}(\alpha_i)
\nonumber \\
&&\langle \gamma(q) \vert \bar q(x) g_s G_{\mu \nu}(v x) i
\gamma_\alpha q(0) \vert 0 \rangle = e_q f_{3 \gamma} q_\alpha
(\varepsilon_\mu q_\nu - \varepsilon_\nu q_\mu) \int {\cal
D}\alpha_i e^{i (\alpha_{\bar q} + v \alpha_g) q x} {\cal
V}(\alpha_i) \nonumber\\
&& \langle \gamma(q) \vert \bar q(x)
\sigma_{\alpha \beta} g_s G_{\mu \nu}(v x) q(0) \vert 0 \rangle  =
e_q \langle \bar q q \rangle \left\{
        \left[\left(\varepsilon_\mu - q_\mu \frac{\varepsilon x}{q x}\right)\left(g_{\alpha \nu} -
        \frac{1}{qx} (q_\alpha x_\nu + q_\nu x_\alpha)\right) \right. \right. q_\beta
\nonumber \\ && -
         \left(\varepsilon_\mu - q_\mu \frac{\varepsilon x}{q x}\right)\left(g_{\beta \nu} -
        \frac{1}{qx} (q_\beta x_\nu + q_\nu x_\beta)\right) q_\alpha
-
         \left(\varepsilon_\nu - q_\nu \frac{\varepsilon x}{q x}\right)\left(g_{\alpha \mu} -
        \frac{1}{qx} (q_\alpha x_\mu + q_\mu x_\alpha)\right) q_\beta
\nonumber \\ &&+
         \left. \left(\varepsilon_\nu - q_\nu \frac{\varepsilon x}{q.x}\right)\left( g_{\beta \mu} -
        \frac{1}{qx} (q_\beta x_\mu + q_\mu x_\beta)\right) q_\alpha \right]
   \int {\cal D}\alpha_i e^{i (\alpha_{\bar q} + v \alpha_g) qx} {\cal T}_1(\alpha_i)
\nonumber \\ &&+
        \left[\left(\varepsilon_\alpha - q_\alpha \frac{\varepsilon x}{qx}\right)
        \left(g_{\mu \beta} - \frac{1}{qx}(q_\mu x_\beta + q_\beta x_\mu)\right) \right. q_\nu
\nonumber \\ &&-
         \left(\varepsilon_\alpha - q_\alpha \frac{\varepsilon x}{qx}\right)
        \left(g_{\nu \beta} - \frac{1}{qx}(q_\nu x_\beta + q_\beta x_\nu)\right)  q_\mu
\nonumber \\ && -
         \left(\varepsilon_\beta - q_\beta \frac{\varepsilon x}{qx}\right)
        \left(g_{\mu \alpha} - \frac{1}{qx}(q_\mu x_\alpha + q_\alpha x_\mu)\right) q_\nu
\nonumber \\ &&+
         \left. \left(\varepsilon_\beta - q_\beta \frac{\varepsilon x}{qx}\right)
        \left(g_{\nu \alpha} - \frac{1}{qx}(q_\nu x_\alpha + q_\alpha x_\nu) \right) q_\mu
        \right]
    \int {\cal D} \alpha_i e^{i (\alpha_{\bar q} + v \alpha_g) qx} {\cal T}_2(\alpha_i)
\nonumber \\
&&+\frac{1}{qx} (q_\mu x_\nu - q_\nu x_\mu)
        (\varepsilon_\alpha q_\beta - \varepsilon_\beta q_\alpha)
    \int {\cal D} \alpha_i e^{i (\alpha_{\bar q} + v \alpha_g) qx} {\cal T}_3(\alpha_i)
\nonumber \\ &&+
        \left. \frac{1}{qx} (q_\alpha x_\beta - q_\beta x_\alpha)
        (\varepsilon_\mu q_\nu - \varepsilon_\nu q_\mu)
    \int {\cal D} \alpha_i e^{i (\alpha_{\bar q} + v \alpha_g) qx} {\cal T}_4(\alpha_i)
                        \right\}~,
\end{eqnarray*}
where $\varphi_\gamma(u)$ is the leading twist-2, $\psi^v(u)$,
$\psi^a(u)$, ${\cal A}(\alpha_i)$ and ${\cal V}(\alpha_i)$, are the twist-3, and
$h_\gamma(u)$, $\mathbb{A}(u)$, ${\cal S}(\alpha_i)$, ${\cal{\tilde S}}(\alpha_i)$, ${\cal T}_1(\alpha_i)$, ${\cal T}_2(\alpha_i)$, ${\cal T}_3(\alpha_i)$
and ${\cal T}_4(\alpha_i)$ are the
twist-4 photon DAs.
The measure ${\cal D} \alpha_i$ is defined as
\begin{eqnarray*}
\label{nolabel05}
\int {\cal D} \alpha_i = \int_0^1 d \alpha_{\bar q} \int_0^1 d
\alpha_q \int_0^1 d \alpha_g \delta(1-\alpha_{\bar
q}-\alpha_q-\alpha_g)~.\nonumber
\end{eqnarray*}

The expressions of the DAs entering into the above matrix elements are
defined as:

\begin{eqnarray*}
\varphi_\gamma(u) &=& 6 u \bar u \left( 1 + \varphi_2(\mu)
C_2^{\frac{3}{2}}(u - \bar u) \right),
\nonumber \\
\psi^v(u) &=& 3 \left(3 (2 u - 1)^2 -1 \right)+\frac{3}{64} \left(15
w^V_\gamma - 5 w^A_\gamma\right)
                        \left(3 - 30 (2 u - 1)^2 + 35 (2 u -1)^4
                        \right),
\nonumber \\
\psi^a(u) &=& \left(1- (2 u -1)^2\right)\left(5 (2 u -1)^2 -1\right)
\frac{5}{2}
    \left(1 + \frac{9}{16} w^V_\gamma - \frac{3}{16} w^A_\gamma
    \right),
\nonumber \\
h_\gamma(u) &=& - 10 \left(1 + 2 \kappa^+\right) C_2^{\frac{1}{2}}(u
- \bar u),
\nonumber \\
\mathbb{A}(u) &=& 40 u^2 \bar u^2 \left(3 \kappa - \kappa^+
+1\right)  +
        8 (\zeta_2^+ - 3 \zeta_2) \left[u \bar u (2 + 13 u \bar u) \right.
\nonumber \\ && + \left.
                2 u^3 (10 -15 u + 6 u^2) \ln(u) + 2 \bar u^3 (10 - 15 \bar u + 6 \bar u^2)
        \ln(\bar u) \right],
\nonumber \\
{\cal A}(\alpha_i) &=& 360 \alpha_q \alpha_{\bar q} \alpha_g^2
        \left(1 + w^A_\gamma \frac{1}{2} (7 \alpha_g - 3)\right),
\nonumber \\
{\cal V}(\alpha_i) &=& 540 w^V_\gamma (\alpha_q - \alpha_{\bar q})
\alpha_q \alpha_{\bar q}
                \alpha_g^2,
\nonumber \\
{\cal T}_1(\alpha_i) &=& -120 (3 \zeta_2 + \zeta_2^+)(\alpha_{\bar
q} - \alpha_q)
        \alpha_{\bar q} \alpha_q \alpha_g,
\nonumber \\
{\cal T}_2(\alpha_i) &=& 30 \alpha_g^2 (\alpha_{\bar q} - \alpha_q)
    \left((\kappa - \kappa^+) + (\zeta_1 - \zeta_1^+)(1 - 2\alpha_g) +
    \zeta_2 (3 - 4 \alpha_g)\right),
\nonumber \\
{\cal T}_3(\alpha_i) &=& - 120 (3 \zeta_2 - \zeta_2^+)(\alpha_{\bar
q} -\alpha_q)
        \alpha_{\bar q} \alpha_q \alpha_g,
\nonumber \\
{\cal T}_4(\alpha_i) &=& 30 \alpha_g^2 (\alpha_{\bar q} - \alpha_q)
    \left((\kappa + \kappa^+) + (\zeta_1 + \zeta_1^+)(1 - 2\alpha_g) +
    \zeta_2 (3 - 4 \alpha_g)\right),\nonumber \\
{\cal S}(\alpha_i) &=& 30\alpha_g^2\{(\kappa +
\kappa^+)(1-\alpha_g)+(\zeta_1 + \zeta_1^+)(1 - \alpha_g)(1 -
2\alpha_g)\nonumber +\zeta_2[3 (\alpha_{\bar q} - \alpha_q)^2-\alpha_g(1 - \alpha_g)]\},\nonumber \\
\tilde {\cal S}(\alpha_i) &=&-30\alpha_g^2\{(\kappa -\kappa^+)(1-\alpha_g)+(\zeta_1 - \zeta_1^+)(1 - \alpha_g)(1 -
2\alpha_g)\nonumber +\zeta_2 [3 (\alpha_{\bar q} -\alpha_q)^2-\alpha_g(1 - \alpha_g)]\}.
\end{eqnarray*}
Numerical values of parameters used in DAs; $\varphi_2(1~GeV) = 0$, 
$w^V_\gamma = 3.8 \pm 1.8$, $w^A_\gamma = -2.1 \pm 1.0$, 
$\kappa = 0.2$, $\kappa^+ = 0$, $\zeta_1 = 0.4$, $\zeta_2 = 0.3$, 
$\zeta_1^+ = 0$, and $\zeta_2^+ = 0$.

\bibliography{refs}

\begin{thebibliography}{10}
\expandafter\ifx\csname url\endcsname\relax
  \def\url#1{\texttt{#1}}\fi
\expandafter\ifx\csname urlprefix\endcsname\relax\def\urlprefix{URL }\fi
\expandafter\ifx\csname href\endcsname\relax
  \def\href#1#2{#2} \def\path#1{#1}\fi

\bibitem{Mattson:2002vu}
M.~Mattson, et~al., {First observation of the doubly charmed baryon
  $\Xi^+_{cc}$}, Phys. Rev. Lett. 89 (2002) 112001.
\newblock \href {http://arxiv.org/abs/hep-ex/0208014}
  {\path{arXiv:hep-ex/0208014}}, \href
  {https://doi.org/10.1103/PhysRevLett.89.112001}
  {\path{doi:10.1103/PhysRevLett.89.112001}}.

\bibitem{Chistov:2006zj}
R.~Chistov, et~al., {Observation of new states decaying into $\Lambda_c^+ K^-
  \pi^+$ and $\Lambda_c^+ K^0_s \pi^-$}, Phys. Rev. Lett. 97 (2006) 162001.
\newblock \href {http://arxiv.org/abs/hep-ex/0606051}
  {\path{arXiv:hep-ex/0606051}}, \href
  {https://doi.org/10.1103/PhysRevLett.97.162001}
  {\path{doi:10.1103/PhysRevLett.97.162001}}.

\bibitem{Ratti:2003ez}
S.~P. Ratti, {New results on c-baryons and a search for cc-baryons in FOCUS},
  Nucl. Phys. Proc. Suppl. 115 (2003) 33--36, [,33(2003)].
\newblock \href {https://doi.org/10.1016/S0920-5632(02)01948-5}
  {\path{doi:10.1016/S0920-5632(02)01948-5}}.

\bibitem{Aubert:2006qw}
B.~Aubert, et~al., {Search for doubly charmed baryons $\Xi_{cc}^+$ and
  $\Xi_{cc}^{++}$ in BABAR}, Phys. Rev. D74 (2006) 011103.
\newblock \href {http://arxiv.org/abs/hep-ex/0605075}
  {\path{arXiv:hep-ex/0605075}}, \href
  {https://doi.org/10.1103/PhysRevD.74.011103}
  {\path{doi:10.1103/PhysRevD.74.011103}}.

\bibitem{Aaij:2017ueg}
R.~Aaij, et~al., {Observation of the doubly charmed baryon $\Xi_{cc}^{++}$},
  Phys. Rev. Lett. 119~(11) (2017) 112001.
\newblock \href {http://arxiv.org/abs/1707.01621} {\path{arXiv:1707.01621}},
  \href {https://doi.org/10.1103/PhysRevLett.119.112001}
  {\path{doi:10.1103/PhysRevLett.119.112001}}.

\bibitem{Bagan:1992za}
E.~Bagan, M.~Chabab, S.~Narison, {Baryons with two heavy quarks from QCD
  spectral sum rules}, Phys. Lett. B306 (1993) 350--356.
\newblock \href {https://doi.org/10.1016/0370-2693(93)90090-5}
  {\path{doi:10.1016/0370-2693(93)90090-5}}.

\bibitem{Roncaglia:1995az}
R.~Roncaglia, D.~B. Lichtenberg, E.~Predazzi, {Predicting the masses of baryons
  containing one or two heavy quarks}, Phys. Rev. D52 (1995) 1722--1725.
\newblock \href {http://arxiv.org/abs/hep-ph/9502251}
  {\path{arXiv:hep-ph/9502251}}, \href
  {https://doi.org/10.1103/PhysRevD.52.1722}
  {\path{doi:10.1103/PhysRevD.52.1722}}.

\bibitem{Ebert:1996ec}
D.~Ebert, R.~N. Faustov, V.~O. Galkin, A.~P. Martynenko, V.~A. Saleev, {Heavy
  baryons in the relativistic quark model}, Z. Phys. C76 (1997) 111--115.
\newblock \href {http://arxiv.org/abs/hep-ph/9607314}
  {\path{arXiv:hep-ph/9607314}}, \href {https://doi.org/10.1007/s002880050534}
  {\path{doi:10.1007/s002880050534}}.

\bibitem{Tong:1999qs}
S.-P. Tong, Y.-B. Ding, X.-H. Guo, H.-Y. Jin, X.-Q. Li, P.-N. Shen, R.~Zhang,
  {Spectra of baryons containing two heavy quarks in potential model}, Phys.
  Rev. D62 (2000) 054024.
\newblock \href {http://arxiv.org/abs/hep-ph/9910259}
  {\path{arXiv:hep-ph/9910259}}, \href
  {https://doi.org/10.1103/PhysRevD.62.054024}
  {\path{doi:10.1103/PhysRevD.62.054024}}.

\bibitem{Itoh:2000um}
C.~Itoh, T.~Minamikawa, K.~Miura, T.~Watanabe, {Doubly charmed baryon masses
  and quark wave functions in baryons}, Phys. Rev. D61 (2000) 057502.
\newblock \href {https://doi.org/10.1103/PhysRevD.61.057502}
  {\path{doi:10.1103/PhysRevD.61.057502}}.

\bibitem{Gershtein:2000nx}
S.~S. Gershtein, V.~V. Kiselev, A.~K. Likhoded, A.~I. Onishchenko,
  {Spectroscopy of doubly heavy baryons}, Phys. Rev. D62 (2000) 054021.
\newblock \href {https://doi.org/10.1103/PhysRevD.62.054021}
  {\path{doi:10.1103/PhysRevD.62.054021}}.

\bibitem{Kiselev:2001fw}
V.~V. Kiselev, A.~K. Likhoded, {Baryons with two heavy quarks}, Phys. Usp. 45
  (2002) 455--506.
\newblock \href {http://arxiv.org/abs/hep-ph/0103169}
  {\path{arXiv:hep-ph/0103169}}, \href
  {https://doi.org/10.1070/PU2002v045n05ABEH000958}
  {\path{doi:10.1070/PU2002v045n05ABEH000958}}.

\bibitem{Kiselev:2002iy}
V.~V. Kiselev, A.~K. Likhoded, O.~N. Pakhomova, V.~A. Saleev, {Mass spectra of
  doubly heavy Omega $Q Q^\prime$ baryons}, Phys. Rev. D66 (2002) 034030.
\newblock \href {http://arxiv.org/abs/hep-ph/0206140}
  {\path{arXiv:hep-ph/0206140}}, \href
  {https://doi.org/10.1103/PhysRevD.66.034030}
  {\path{doi:10.1103/PhysRevD.66.034030}}.

\bibitem{Narodetskii:2001bq}
I.~M. Narodetskii, M.~A. Trusov, {The Heavy baryons in the nonperturbative
  string approach}, Phys. Atom. Nucl. 65 (2002) 917--924, [Yad.
  Fiz.65,949(2002)].
\newblock \href {http://arxiv.org/abs/hep-ph/0104019}
  {\path{arXiv:hep-ph/0104019}}, \href {https://doi.org/10.1134/1.1481486}
  {\path{doi:10.1134/1.1481486}}.

\bibitem{Lewis:2001iz}
R.~Lewis, N.~Mathur, R.~M. Woloshyn, {Charmed baryons in lattice QCD}, Phys.
  Rev. D64 (2001) 094509.
\newblock \href {http://arxiv.org/abs/hep-ph/0107037}
  {\path{arXiv:hep-ph/0107037}}, \href
  {https://doi.org/10.1103/PhysRevD.64.094509}
  {\path{doi:10.1103/PhysRevD.64.094509}}.

\bibitem{Ebert:2002ig}
D.~Ebert, R.~N. Faustov, V.~O. Galkin, A.~P. Martynenko, {Mass spectra of
  doubly heavy baryons in the relativistic quark model}, Phys. Rev. D66 (2002)
  014008.
\newblock \href {http://arxiv.org/abs/hep-ph/0201217}
  {\path{arXiv:hep-ph/0201217}}, \href
  {https://doi.org/10.1103/PhysRevD.66.014008}
  {\path{doi:10.1103/PhysRevD.66.014008}}.

\bibitem{Mathur:2002ce}
N.~Mathur, R.~Lewis, R.~M. Woloshyn, {Charmed and bottom baryons from lattice
  NRQCD}, Phys. Rev. D66 (2002) 014502.
\newblock \href {http://arxiv.org/abs/hep-ph/0203253}
  {\path{arXiv:hep-ph/0203253}}, \href
  {https://doi.org/10.1103/PhysRevD.66.014502}
  {\path{doi:10.1103/PhysRevD.66.014502}}.

\bibitem{Flynn:2003vz}
J.~M. Flynn, F.~Mescia, A.~S.~B. Tariq, {Spectroscopy of doubly charmed baryons
  in lattice QCD}, JHEP 07 (2003) 066.
\newblock \href {http://arxiv.org/abs/hep-lat/0307025}
  {\path{arXiv:hep-lat/0307025}}, \href
  {https://doi.org/10.1088/1126-6708/2003/07/066}
  {\path{doi:10.1088/1126-6708/2003/07/066}}.

\bibitem{Vijande:2004at}
J.~Vijande, H.~Garcilazo, A.~Valcarce, F.~Fernandez, {Spectroscopy of doubly
  charmed baryons}, Phys. Rev. D70 (2004) 054022.
\newblock \href {http://arxiv.org/abs/hep-ph/0408274}
  {\path{arXiv:hep-ph/0408274}}, \href
  {https://doi.org/10.1103/PhysRevD.70.054022}
  {\path{doi:10.1103/PhysRevD.70.054022}}.

\bibitem{Chiu:2005zc}
T.-W. Chiu, T.-H. Hsieh, {Baryon masses in lattice QCD with exact chiral
  symmetry}, Nucl. Phys. A755 (2005) 471--474.
\newblock \href {http://arxiv.org/abs/hep-lat/0501021}
  {\path{arXiv:hep-lat/0501021}}, \href
  {https://doi.org/10.1016/j.nuclphysa.2005.03.090}
  {\path{doi:10.1016/j.nuclphysa.2005.03.090}}.

\bibitem{Migura:2006ep}
S.~Migura, D.~Merten, B.~Metsch, H.-R. Petry, {Charmed baryons in a
  relativistic quark model}, Eur. Phys. J. A28 (2006) 41.
\newblock \href {http://arxiv.org/abs/hep-ph/0602153}
  {\path{arXiv:hep-ph/0602153}}, \href
  {https://doi.org/10.1140/epja/i2006-10017-9}
  {\path{doi:10.1140/epja/i2006-10017-9}}.

\bibitem{Albertus:2006ya}
C.~Albertus, E.~Hernandez, J.~Nieves, J.~M. Verde-Velasco, {Static properties
  and semileptonic decays of doubly heavy baryons in a nonrelativistic quark
  model}, Eur. Phys. J. A32 (2007) 183--199, [Erratum: Eur. Phys.
  J.A36,119(2008)].
\newblock \href {http://arxiv.org/abs/hep-ph/0610030}
  {\path{arXiv:hep-ph/0610030}}, \href
  {https://doi.org/10.1140/epja/i2007-10364-y; 10.1140/epja/i2008-10547-0}
  {\path{doi:10.1140/epja/i2007-10364-y; 10.1140/epja/i2008-10547-0}}.

\bibitem{Martynenko:2007je}
A.~P. Martynenko, {Ground-state triply and doubly heavy baryons in a
  relativistic three-quark model}, Phys. Lett. B663 (2008) 317--321.
\newblock \href {http://arxiv.org/abs/0708.2033} {\path{arXiv:0708.2033}},
  \href {https://doi.org/10.1016/j.physletb.2008.04.030}
  {\path{doi:10.1016/j.physletb.2008.04.030}}.

\bibitem{Tang:2011fv}
L.~Tang, X.-H. Yuan, C.-F. Qiao, X.-Q. Li, {Study of Doubly Heavy Baryon
  Spectrum via QCD Sum Rules}, Commun. Theor. Phys. 57 (2012) 435--444.
\newblock \href {http://arxiv.org/abs/1104.4934} {\path{arXiv:1104.4934}},
  \href {https://doi.org/10.1088/0253-6102/57/3/15}
  {\path{doi:10.1088/0253-6102/57/3/15}}.

\bibitem{Liu:2007fg}
X.~Liu, H.-X. Chen, Y.-R. Liu, A.~Hosaka, S.-L. Zhu, {Bottom baryons}, Phys.
  Rev. D77 (2008) 014031.
\newblock \href {http://arxiv.org/abs/0710.0123} {\path{arXiv:0710.0123}},
  \href {https://doi.org/10.1103/PhysRevD.77.014031}
  {\path{doi:10.1103/PhysRevD.77.014031}}.

\bibitem{Roberts:2007ni}
W.~Roberts, M.~Pervin, {Heavy baryons in a quark model}, Int. J. Mod. Phys. A23
  (2008) 2817--2860.
\newblock \href {http://arxiv.org/abs/0711.2492} {\path{arXiv:0711.2492}},
  \href {https://doi.org/10.1142/S0217751X08041219}
  {\path{doi:10.1142/S0217751X08041219}}.

\bibitem{Valcarce:2008dr}
A.~Valcarce, H.~Garcilazo, J.~Vijande, {Towards an understanding of heavy
  baryon spectroscopy}, Eur. Phys. J. A37 (2008) 217--225.
\newblock \href {http://arxiv.org/abs/0807.2973} {\path{arXiv:0807.2973}},
  \href {https://doi.org/10.1140/epja/i2008-10616-4}
  {\path{doi:10.1140/epja/i2008-10616-4}}.

\bibitem{Liu:2009jc}
L.~Liu, H.-W. Lin, K.~Orginos, A.~Walker-Loud, {Singly and Doubly Charmed J=1/2
  Baryon Spectrum from Lattice QCD}, Phys. Rev. D81 (2010) 094505.
\newblock \href {http://arxiv.org/abs/0909.3294} {\path{arXiv:0909.3294}},
  \href {https://doi.org/10.1103/PhysRevD.81.094505}
  {\path{doi:10.1103/PhysRevD.81.094505}}.

\bibitem{Alexandrou:2012xk}
C.~Alexandrou, J.~Carbonell, D.~Christaras, V.~Drach, M.~Gravina, M.~Papinutto,
  {Strange and charm baryon masses with two flavors of dynamical twisted mass
  fermions}, Phys. Rev. D86 (2012) 114501.
\newblock \href {http://arxiv.org/abs/1205.6856} {\path{arXiv:1205.6856}},
  \href {https://doi.org/10.1103/PhysRevD.86.114501}
  {\path{doi:10.1103/PhysRevD.86.114501}}.

\bibitem{Aliev:2012ru}
T.~M. Aliev, K.~Azizi, M.~Savci, {Doubly Heavy Spin--1/2 Baryon Spectrum in
  QCD}, Nucl. Phys. A895 (2012) 59--70.
\newblock \href {http://arxiv.org/abs/1205.2873} {\path{arXiv:1205.2873}},
  \href {https://doi.org/10.1016/j.nuclphysa.2012.09.009}
  {\path{doi:10.1016/j.nuclphysa.2012.09.009}}.

\bibitem{Aliev:2012iv}
T.~M. Aliev, K.~Azizi, M.~Savci, {The masses and residues of doubly heavy
  spin-3/2 baryons}, J. Phys. G40 (2013) 065003.
\newblock \href {http://arxiv.org/abs/1208.1976} {\path{arXiv:1208.1976}},
  \href {https://doi.org/10.1088/0954-3899/40/6/065003}
  {\path{doi:10.1088/0954-3899/40/6/065003}}.

\bibitem{Namekawa:2013vu}
Y.~Namekawa, et~al., {Charmed baryons at the physical point in 2+1 flavor
  lattice QCD}, Phys. Rev. D87~(9) (2013) 094512.
\newblock \href {http://arxiv.org/abs/1301.4743} {\path{arXiv:1301.4743}},
  \href {https://doi.org/10.1103/PhysRevD.87.094512}
  {\path{doi:10.1103/PhysRevD.87.094512}}.

\bibitem{Karliner:2014gca}
M.~Karliner, J.~L. Rosner, {Baryons with two heavy quarks: Masses, production,
  decays, and detection}, Phys. Rev. D90~(9) (2014) 094007.
\newblock \href {http://arxiv.org/abs/1408.5877} {\path{arXiv:1408.5877}},
  \href {https://doi.org/10.1103/PhysRevD.90.094007}
  {\path{doi:10.1103/PhysRevD.90.094007}}.

\bibitem{Sun:2014aya}
Z.-F. Sun, Z.-W. Liu, X.~Liu, S.-L. Zhu, {Masses and axial currents of the
  doubly charmed baryons}, Phys. Rev. D91~(9) (2015) 094030.
\newblock \href {http://arxiv.org/abs/1411.2117} {\path{arXiv:1411.2117}},
  \href {https://doi.org/10.1103/PhysRevD.91.094030}
  {\path{doi:10.1103/PhysRevD.91.094030}}.

\bibitem{Chen:2015kpa}
H.-X. Chen, W.~Chen, Q.~Mao, A.~Hosaka, X.~Liu, S.-L. Zhu, {P-wave charmed
  baryons from QCD sum rules}, Phys. Rev. D91~(5) (2015) 054034.
\newblock \href {http://arxiv.org/abs/1502.01103} {\path{arXiv:1502.01103}},
  \href {https://doi.org/10.1103/PhysRevD.91.054034}
  {\path{doi:10.1103/PhysRevD.91.054034}}.

\bibitem{Sun:2016wzh}
Z.-F. Sun, M.~J. {Vicente Vacas}, {Masses of doubly charmed baryons in the
  extended on-mass-shell renormalization scheme}, Phys. Rev. D93~(9) (2016)
  094002.
\newblock \href {http://arxiv.org/abs/1602.04714} {\path{arXiv:1602.04714}},
  \href {https://doi.org/10.1103/PhysRevD.93.094002}
  {\path{doi:10.1103/PhysRevD.93.094002}}.

\bibitem{Shah:2016vmd}
Z.~Shah, K.~Thakkar, A.~K. Rai, {Excited State Mass spectra of doubly heavy
  baryons $\Omega_{cc}$, $\Omega_{bb}$ and $\Omega_{bc}$}, Eur. Phys. J.
  C76~(10) (2016) 530.
\newblock \href {http://arxiv.org/abs/1609.03030} {\path{arXiv:1609.03030}},
  \href {https://doi.org/10.1140/epjc/s10052-016-4379-z}
  {\path{doi:10.1140/epjc/s10052-016-4379-z}}.

\bibitem{Kiselev:2017eic}
V.~V. Kiselev, A.~V. Berezhnoy, A.~K. Likhoded, {Quark--Diquark Structure and
  Masses of Doubly Charmed Baryons}, Phys. Atom. Nucl. 81~(3) (2018) 369--372,
  [Yad. Fiz.81,no.3,356(2018)].
\newblock \href {http://arxiv.org/abs/1706.09181} {\path{arXiv:1706.09181}},
  \href {https://doi.org/10.1134/S1063778818030134}
  {\path{doi:10.1134/S1063778818030134}}.

\bibitem{Chen:2017sbg}
H.-X. Chen, Q.~Mao, W.~Chen, X.~Liu, S.-L. Zhu, {Establishing low-lying doubly
  charmed baryons}, Phys. Rev. D96~(3) (2017) 031501, [Erratum: Phys.
  Rev.D96,no.11,119902(2017)].
\newblock \href {http://arxiv.org/abs/1707.01779} {\path{arXiv:1707.01779}},
  \href {https://doi.org/10.1103/PhysRevD.96.031501;
  10.1103/PhysRevD.96.119902} {\path{doi:10.1103/PhysRevD.96.031501;
  10.1103/PhysRevD.96.119902}}.

\bibitem{Hu:2005gf}
J.~Hu, T.~Mehen, {Chiral Lagrangian with heavy quark-diquark symmetry}, Phys.
  Rev. D73 (2006) 054003.
\newblock \href {http://arxiv.org/abs/hep-ph/0511321}
  {\path{arXiv:hep-ph/0511321}}, \href
  {https://doi.org/10.1103/PhysRevD.73.054003}
  {\path{doi:10.1103/PhysRevD.73.054003}}.

\bibitem{Meng:2017fwb}
L.~Meng, N.~Li, S.-L. Zhu, {Deuteron-like states composed of two doubly charmed
  baryons}, Phys. Rev. D95~(11) (2017) 114019.
\newblock \href {http://arxiv.org/abs/1704.01009} {\path{arXiv:1704.01009}},
  \href {https://doi.org/10.1103/PhysRevD.95.114019}
  {\path{doi:10.1103/PhysRevD.95.114019}}.

\bibitem{Narison:2010py}
S.~Narison, R.~Albuquerque, {Mass-splittings of doubly heavy baryons in QCD},
  Phys. Lett. B694 (2011) 217--225.
\newblock \href {http://arxiv.org/abs/1006.2091} {\path{arXiv:1006.2091}},
  \href {https://doi.org/10.1016/j.physletb.2010.09.051}
  {\path{doi:10.1016/j.physletb.2010.09.051}}.

\bibitem{Zhang:2008rt}
J.-R. Zhang, M.-Q. Huang, {Doubly heavy baryons in QCD sum rules}, Phys. Rev.
  D78 (2008) 094007.
\newblock \href {http://arxiv.org/abs/0810.5396} {\path{arXiv:0810.5396}},
  \href {https://doi.org/10.1103/PhysRevD.78.094007}
  {\path{doi:10.1103/PhysRevD.78.094007}}.

\bibitem{Guo:2017vcf}
Z.-H. Guo, {Prediction of exotic doubly charmed baryons within chiral effective
  field theory}, Phys. Rev. D96~(7) (2017) 074004.
\newblock \href {http://arxiv.org/abs/1708.04145} {\path{arXiv:1708.04145}},
  \href {https://doi.org/10.1103/PhysRevD.96.074004}
  {\path{doi:10.1103/PhysRevD.96.074004}}.

\bibitem{Lu:2017meb}
Q.-F. L{\"u}, K.-L. Wang, L.-Y. Xiao, X.-H. Zhong, {Mass spectra and radiative
  transitions of doubly heavy baryons in a relativized quark model}, Phys. Rev.
  D96~(11) (2017) 114006.
\newblock \href {http://arxiv.org/abs/1708.04468} {\path{arXiv:1708.04468}},
  \href {https://doi.org/10.1103/PhysRevD.96.114006}
  {\path{doi:10.1103/PhysRevD.96.114006}}.

\bibitem{Xiao:2017udy}
L.-Y. Xiao, K.-L. Wang, Q.-f. Lu, X.-H. Zhong, S.-L. Zhu, {Strong and radiative
  decays of the doubly charmed baryons}, Phys. Rev. D96~(9) (2017) 094005.
\newblock \href {http://arxiv.org/abs/1708.04384} {\path{arXiv:1708.04384}},
  \href {https://doi.org/10.1103/PhysRevD.96.094005}
  {\path{doi:10.1103/PhysRevD.96.094005}}.

\bibitem{Weng:2018mmf}
X.-Z. Weng, X.-L. Chen, W.-Z. Deng, {Masses of doubly heavy-quark baryons in an
  extended chromomagnetic model}, Phys. Rev. D97~(5) (2018) 054008.
\newblock \href {http://arxiv.org/abs/1801.08644} {\path{arXiv:1801.08644}},
  \href {https://doi.org/10.1103/PhysRevD.97.054008}
  {\path{doi:10.1103/PhysRevD.97.054008}}.

\bibitem{Can:2013zpa}
K.~U. Can, G.~Erkol, B.~Isildak, M.~Oka, T.~T. Takahashi, {Electromagnetic
  properties of doubly charmed baryons in Lattice QCD}, Phys. Lett. B726 (2013)
  703--709.
\newblock \href {http://arxiv.org/abs/1306.0731} {\path{arXiv:1306.0731}},
  \href {https://doi.org/10.1016/j.physletb.2013.09.024}
  {\path{doi:10.1016/j.physletb.2013.09.024}}.

\bibitem{Branz:2010pq}
T.~Branz, A.~Faessler, T.~Gutsche, M.~A. Ivanov, J.~G. Korner, V.~E.
  Lyubovitskij, B.~Oexl, {Radiative decays of double heavy baryons in a
  relativistic constituent three--quark model including hyperfine mixing},
  Phys. Rev. D81 (2010) 114036.
\newblock \href {http://arxiv.org/abs/1005.1850} {\path{arXiv:1005.1850}},
  \href {https://doi.org/10.1103/PhysRevD.81.114036}
  {\path{doi:10.1103/PhysRevD.81.114036}}.

\bibitem{Bose:1980vy}
S.~K. Bose, L.~P. Singh, {Magnetic Moments of Charmed and $B$ Flavored Hadrons
  in {MIT} Bag Model}, Phys. Rev. D22 (1980) 773.
\newblock \href {https://doi.org/10.1103/PhysRevD.22.773}
  {\path{doi:10.1103/PhysRevD.22.773}}.

\bibitem{Patel:2008xs}
B.~Patel, A.~K. Rai, P.~C. Vinodkumar, {Masses and Magnetic Moments of Charmed
  Baryons Using Hyper Central Model}, in: {Proceedings, 12th International
  Conference on Hadron spectroscopy (Hadron 2007): Frascati, Italy, October
  7-13, 2007}, 2008.
\newblock \href {http://arxiv.org/abs/0803.0221} {\path{arXiv:0803.0221}}.

\bibitem{SilvestreBrac:1996bg}
B.~Silvestre-Brac, {Spectrum and static properties of heavy baryons}, Few Body
  Syst. 20 (1996) 1--25.
\newblock \href {https://doi.org/10.1007/s006010050028}
  {\path{doi:10.1007/s006010050028}}.

\bibitem{Patel:2007gx}
B.~Patel, A.~K. Rai, P.~C. Vinodkumar, {Masses and magnetic moments of heavy
  flavour baryons in hyper central model}, J. Phys. G35 (2008) 065001, [J.
  Phys. Conf. Ser.110,122010(2008)].
\newblock \href {http://arxiv.org/abs/0710.3828} {\path{arXiv:0710.3828}},
  \href {https://doi.org/10.1088/1742-6596/110/12/122010;
  10.1088/0954-3899/35/6/065001} {\path{doi:10.1088/1742-6596/110/12/122010;
  10.1088/0954-3899/35/6/065001}}.

\bibitem{Gadaria:2016omw}
A.~N. Gadaria, N.~R. Soni, J.~N. Pandya, {Masses and magnetic moment of doubly
  heavy baryons}, DAE Symp. Nucl. Phys. 61 (2016) 698--699.

\bibitem{JuliaDiaz:2004vh}
B.~Julia-Diaz, D.~O. Riska, {Baryon magnetic moments in relativistic quark
  models}, Nucl. Phys. A739 (2004) 69--88.
\newblock \href {http://arxiv.org/abs/hep-ph/0401096}
  {\path{arXiv:hep-ph/0401096}}, \href
  {https://doi.org/10.1016/j.nuclphysa.2004.03.078}
  {\path{doi:10.1016/j.nuclphysa.2004.03.078}}.

\bibitem{Faessler:2006ft}
A.~Faessler, T.~Gutsche, M.~A. Ivanov, J.~G. Korner, V.~E. Lyubovitskij,
  D.~Nicmorus, K.~Pumsa-ard, {Magnetic moments of heavy baryons in the
  relativistic three-quark model}, Phys. Rev. D73 (2006) 094013.
\newblock \href {http://arxiv.org/abs/hep-ph/0602193}
  {\path{arXiv:hep-ph/0602193}}, \href
  {https://doi.org/10.1103/PhysRevD.73.094013}
  {\path{doi:10.1103/PhysRevD.73.094013}}.

\bibitem{Can:2013tna}
K.~U. Can, G.~Erkol, B.~Isildak, M.~Oka, T.~T. Takahashi, {Electromagnetic
  structure of charmed baryons in Lattice QCD}, JHEP 05 (2014) 125.
\newblock \href {http://arxiv.org/abs/1310.5915} {\path{arXiv:1310.5915}},
  \href {https://doi.org/10.1007/JHEP05(2014)125}
  {\path{doi:10.1007/JHEP05(2014)125}}.

\bibitem{Li:2017cfz}
H.-S. Li, L.~Meng, Z.-W. Liu, S.-L. Zhu, {Magnetic moments of the doubly
  charmed and bottom baryons}, Phys. Rev. D96~(7) (2017) 076011.
\newblock \href {http://arxiv.org/abs/1707.02765} {\path{arXiv:1707.02765}},
  \href {https://doi.org/10.1103/PhysRevD.96.076011}
  {\path{doi:10.1103/PhysRevD.96.076011}}.

\bibitem{Bernotas:2012nz}
A.~Bernotas, V.~Simonis, {Magnetic moments of heavy baryons in the bag model
  reexamined}\href {http://arxiv.org/abs/1209.2900} {\path{arXiv:1209.2900}}.

\bibitem{Lichtenberg:1976fi}
D.~B. Lichtenberg, {Magnetic Moments of Charmed Baryons in the Quark Model},
  Phys. Rev. D15 (1977) 345.
\newblock \href {https://doi.org/10.1103/PhysRevD.15.345}
  {\path{doi:10.1103/PhysRevD.15.345}}.

\bibitem{Oh:1991ws}
Y.-s. Oh, D.-P. Min, M.~Rho, N.~N. Scoccola, {Massive quark baryons as
  skyrmions: Magnetic moments}, Nucl. Phys. A534 (1991) 493--512.
\newblock \href {https://doi.org/10.1016/0375-9474(91)90458-I}
  {\path{doi:10.1016/0375-9474(91)90458-I}}.

\bibitem{Simonis:2018rld}
V.~Simonis, {Improved predictions for magnetic moments and M1 decay widths of
  heavy hadrons}\href {http://arxiv.org/abs/1803.01809}
  {\path{arXiv:1803.01809}}.

\bibitem{Liu:2018euh}
M.-Z. Liu, Y.~Xiao, L.-S. Geng, {Magnetic moments of the spin-1/2 doubly
  charmed baryons in covariant baryon chiral perturbation theory}, Phys. Rev.
  D98~(1) (2018) 014040.
\newblock \href {http://arxiv.org/abs/1807.00912} {\path{arXiv:1807.00912}},
  \href {https://doi.org/10.1103/PhysRevD.98.014040}
  {\path{doi:10.1103/PhysRevD.98.014040}}.

\bibitem{Blin:2018pmj}
A.~N. {Hiller Blin}, Z.-F. Sun, M.~J. {Vicente Vacas}, {Electromagnetic form
  factors of spin 1/2 doubly charmed baryons}, Phys. Rev. D98~(5) (2018)
  054025.
\newblock \href {http://arxiv.org/abs/1807.01059} {\path{arXiv:1807.01059}},
  \href {https://doi.org/10.1103/PhysRevD.98.054025}
  {\path{doi:10.1103/PhysRevD.98.054025}}.

\bibitem{Meng:2017dni}
L.~Meng, H.-S. Li, Z.-W. Liu, S.-L. Zhu, {Magnetic moments of the
  spin-$\frac{3}{2}$ doubly heavy baryons}, Eur. Phys. J. C77~(12) (2017) 869.
\newblock \href {http://arxiv.org/abs/1710.08283} {\path{arXiv:1710.08283}},
  \href {https://doi.org/10.1140/epjc/s10052-017-5447-8}
  {\path{doi:10.1140/epjc/s10052-017-5447-8}}.

\bibitem{Dhir:2009ax}
R.~Dhir, R.~C. Verma, {Magnetic Moments of $J^P = 3/2^+$ Heavy Baryons Using
  Effective Mass Scheme}, Eur. Phys. J. A42 (2009) 243--249.
\newblock \href {http://arxiv.org/abs/0904.2124} {\path{arXiv:0904.2124}},
  \href {https://doi.org/10.1140/epja/i2009-10872-8}
  {\path{doi:10.1140/epja/i2009-10872-8}}.

\bibitem{Li:2017pxa}
H.-S. Li, L.~Meng, Z.-W. Liu, S.-L. Zhu, {Radiative decays of the doubly
  charmed baryons in chiral perturbation theory}, Phys. Lett. B777 (2018)
  169--176.
\newblock \href {http://arxiv.org/abs/1708.03620} {\path{arXiv:1708.03620}},
  \href {https://doi.org/10.1016/j.physletb.2017.12.031}
  {\path{doi:10.1016/j.physletb.2017.12.031}}.

\bibitem{Yu:2017zst}
F.-S. Yu, H.-Y. Jiang, R.-H. Li, C.-D. L{\"u}, W.~Wang, Z.-X. Zhao, {Discovery
  Potentials of Doubly Charmed Baryons}, Chin. Phys. C42~(5) (2018) 051001.
\newblock \href {http://arxiv.org/abs/1703.09086} {\path{arXiv:1703.09086}},
  \href {https://doi.org/10.1088/1674-1137/42/5/051001}
  {\path{doi:10.1088/1674-1137/42/5/051001}}.

\bibitem{Cui:2017udv}
E.-L. Cui, H.-X. Chen, W.~Chen, X.~Liu, S.-L. Zhu, {Suggested search for doubly
  charmed baryons of $J^P=3/2^+$ via their electromagnetic transitions}, Phys.
  Rev. D97~(3) (2018) 034018.
\newblock \href {http://arxiv.org/abs/1712.03615} {\path{arXiv:1712.03615}},
  \href {https://doi.org/10.1103/PhysRevD.97.034018}
  {\path{doi:10.1103/PhysRevD.97.034018}}.

\bibitem{Li:2017ndo}
R.-H. Li, C.-D. L{\"u}, W.~Wang, F.-S. Yu, Z.-T. Zou, {Doubly-heavy baryon weak
  decays: $\Xi_{bc}^{0}\to pK^{-}$ and $\Xi_{cc}^{+}\to
  \Sigma_{c}^{++}(2520)K^{-}$}, Phys. Lett. B767 (2017) 232--235.
\newblock \href {http://arxiv.org/abs/1701.03284} {\path{arXiv:1701.03284}},
  \href {https://doi.org/10.1016/j.physletb.2017.02.003}
  {\path{doi:10.1016/j.physletb.2017.02.003}}.

\bibitem{Wang:2017mqp}
W.~Wang, F.-S. Yu, Z.-X. Zhao, {Weak decays of doubly heavy baryons: the
  $1/2\rightarrow 1/2$ case}, Eur. Phys. J. C77~(11) (2017) 781.
\newblock \href {http://arxiv.org/abs/1707.02834} {\path{arXiv:1707.02834}},
  \href {https://doi.org/10.1140/epjc/s10052-017-5360-1}
  {\path{doi:10.1140/epjc/s10052-017-5360-1}}.

\bibitem{Wang:2017azm}
W.~Wang, Z.-P. Xing, J.~Xu, {Weak Decays of Doubly Heavy Baryons: SU(3)
  Analysis}, Eur. Phys. J. C77~(11) (2017) 800.
\newblock \href {http://arxiv.org/abs/1707.06570} {\path{arXiv:1707.06570}},
  \href {https://doi.org/10.1140/epjc/s10052-017-5363-y}
  {\path{doi:10.1140/epjc/s10052-017-5363-y}}.

\bibitem{Shi:2017dto}
Y.-J. Shi, W.~Wang, Y.~Xing, J.~Xu, {Weak Decays of Doubly Heavy Baryons:
  Multi-body Decay Channels}, Eur. Phys. J. C78~(1) (2018) 56.
\newblock \href {http://arxiv.org/abs/1712.03830} {\path{arXiv:1712.03830}},
  \href {https://doi.org/10.1140/epjc/s10052-018-5532-7}
  {\path{doi:10.1140/epjc/s10052-018-5532-7}}.

\bibitem{Chernyak:1990ag}
V.~L. Chernyak, I.~R. Zhitnitsky, {B meson exclusive decays into baryons},
  Nucl. Phys. B345 (1990) 137--172.
\newblock \href {https://doi.org/10.1016/0550-3213(90)90612-H}
  {\path{doi:10.1016/0550-3213(90)90612-H}}.

\bibitem{Braun:1988qv}
V.~M. Braun, I.~E. Filyanov, {QCD Sum Rules in Exclusive Kinematics and Pion
  Wave Function}, Z. Phys. C44 (1989) 157, [Yad. Fiz.50,818(1989)].
\newblock \href {https://doi.org/10.1007/BF01548594}
  {\path{doi:10.1007/BF01548594}}.

\bibitem{Balitsky:1989ry}
I.~I. Balitsky, V.~M. Braun, A.~V. Kolesnichenko, {Radiative Decay $\Sigma^+
  \rightarrow p \gamma$ in Quantum Chromodynamics}, Nucl. Phys. B312 (1989)
  509--550.
\newblock \href {https://doi.org/10.1016/0550-3213(89)90570-1}
  {\path{doi:10.1016/0550-3213(89)90570-1}}.

\bibitem{Yang:1993bp}
K.-C. Yang, W.~Y.~P. Hwang, E.~M. Henley, L.~S. Kisslinger, {QCD sum rules and
  neutron proton mass difference}, Phys. Rev. D47 (1993) 3001--3012.
\newblock \href {https://doi.org/10.1103/PhysRevD.47.3001}
  {\path{doi:10.1103/PhysRevD.47.3001}}.

\bibitem{Belyaev:1985wza}
V.~M. Belyaev, B.~{\relax Yu}. Blok, {Charmed baryons in Qunatum
  Chromodynamics}, Z. Phys. C30 (1986) 151.
\newblock \href {https://doi.org/10.1007/BF01560689}
  {\path{doi:10.1007/BF01560689}}.

\bibitem{Agaev:2016srl}
S.~S. Agaev, K.~Azizi, H.~Sundu, {Application of the QCD light cone sum rule to
  tetraquarks: the strong vertices $X_bX_b\rho$ and $X_cX_c\rho$}, Phys. Rev.
  D93~(11) (2016) 114036.
\newblock \href {http://arxiv.org/abs/1605.02496} {\path{arXiv:1605.02496}},
  \href {https://doi.org/10.1103/PhysRevD.93.114036}
  {\path{doi:10.1103/PhysRevD.93.114036}}.

\bibitem{Azizi:2018duk}
K.~Azizi, A.~R. Olamaei, S.~Rostami, {Beautiful mathematics for beauty-full and
  other multi-heavy hadronic systems}, Eur. Phys. J. A54~(9) (2018) 162.
\newblock \href {http://arxiv.org/abs/1801.06789} {\path{arXiv:1801.06789}},
  \href {https://doi.org/10.1140/epja/i2018-12595-1}
  {\path{doi:10.1140/epja/i2018-12595-1}}.

\bibitem{Patrignani:2016xqp}
C.~Patrignani, et~al., {Review of Particle Physics}, Chin. Phys. C40~(10)
  (2016) 100001.
\newblock \href {https://doi.org/10.1088/1674-1137/40/10/100001}
  {\path{doi:10.1088/1674-1137/40/10/100001}}.

\bibitem{Ball:2002ps}
P.~Ball, V.~M. Braun, N.~Kivel, {Photon distribution amplitudes in QCD}, Nucl.
  Phys. B649 (2003) 263--296.
\newblock \href {http://arxiv.org/abs/hep-ph/0207307}
  {\path{arXiv:hep-ph/0207307}}, \href
  {https://doi.org/10.1016/S0550-3213(02)01017-9}
  {\path{doi:10.1016/S0550-3213(02)01017-9}}.

\bibitem{Ioffe:2005ym}
B.~L. Ioffe, {QCD at low energies}, Prog. Part. Nucl. Phys. 56 (2006) 232--277.
\newblock \href {http://arxiv.org/abs/hep-ph/0502148}
  {\path{arXiv:hep-ph/0502148}}, \href
  {https://doi.org/10.1016/j.ppnp.2005.05.001}
  {\path{doi:10.1016/j.ppnp.2005.05.001}}.

\bibitem{Rohrwild:2007yt}
J.~Rohrwild, {Determination of the magnetic susceptibility of the quark
  condensate using radiative heavy meson decays}, JHEP 09 (2007) 073.
\newblock \href {http://arxiv.org/abs/0708.1405} {\path{arXiv:0708.1405}},
  \href {https://doi.org/10.1088/1126-6708/2007/09/073}
  {\path{doi:10.1088/1126-6708/2007/09/073}}.

\end{thebibliography}
\end{document}